\documentclass[5p]{elsarticle}
\usepackage{graphicx}

\usepackage{amsmath}
\usepackage{algorithm,algorithmic}
\usepackage{hyperref}
\usepackage{subfig}
\usepackage{multirow}
\usepackage{amssymb}
\usepackage{rotating}

\begin{document}

\begin{frontmatter}
\title{Role-based lateral movement detection with unsupervised learning}

\author{Brian~A.~Powell}
\address{The Johns Hopkins University Applied Physics Laboratory, Laurel, MD 20723}
\ead{brian.a.powell@jhuapl.edu}


\begin{abstract}
Adversarial lateral movement via compromised accounts remains difficult to discover via traditional rule-based defenses because it generally lacks explicit indicators of compromise.  We propose a behavior-based, unsupervised framework comprising two methods of lateral movement detection on enterprise networks: one aimed at generic lateral movement via either exploit or authenticated connections, and one targeting the specific techniques of process injection and hijacking.  The first method is based on the premise that the role of a system---the functions it performs on the network---determines the roles of the systems it should make connections with.  The adversary meanwhile might move between any systems whatever, possibly seeking out systems with unusual roles that facilitate certain accesses.  We use unsupervised learning to cluster systems according to role and identify connections to systems with novel roles as potentially malicious.  The second method is based on the premise that the temporal patterns of inter-system processes that facilitate these connections depend on the roles of the systems involved. If a process is compromised by an attacker, these normal patterns might be disrupted in discernible ways. We apply frequent-itemset mining to process sequences to establish regular patterns of communication between systems based on role, and identify rare process sequences as signalling potentially malicious connections.  
\end{abstract}

\begin{keyword}
intrusion detection, lateral movement
\end{keyword}

\end{frontmatter}

\section{Introduction}
Lateral movement remains one of the most challenging steps of the cyber attack life cycle to detect.  Following a successful intrusion, the adversary will move within the target network, from system-to-system, performing reconnaissance, stealing credentials, and escalating privileges.  Activities conducted via compromised accounts are difficult to detect because they tend to appear normal: accesses are properly authenticated, and authorized software available on the compromised system can be used to further access. Rule-based intrusion detection systems struggle greatly to detect activities like these that lack recognizable signatures of compromise.

Machine learning, and artificial intelligence more broad- ly, offers new avenues of cyber attack detection.  While an overwhelming amount of research has been conducted on applying machine learning to the detection of traditional classes of intrusion (malware propagation, denial of service, botnets, and others), there remains an urgent need to develop capabilities more in line with the techniques of modern advanced threats: namely, the use of authorized services and compromised accounts to move quietly through the network.  While most classes of intrusion have common ``tells'' ({\it e.g.} many connections in quick succession indicate worms or reconnaissance, high volume data transfers indicate exfiltration), authorized lateral movement has no such explicit indicators.  But, if we can bring methods of statistical pattern recognition and behavioral modeling to bear on data describing normal user, system, and network characteristics, it becomes possible to apply unsupervised or one-class learning to discover malicious activities without explicit indicators or historical precedent in the enterprise environment. 

In this paper, we propose two uses of unsupervised learning to discover anomalous connections among systems on a network.  We focus on general connections---not only authentications---and our method is not reliant on any explicit indicators of compromise, artifacts related to adversary tactics or techniques, or the use of any particular protocol or service.   The two approaches are related and fit into a framework that emphasizes that a system's {\it role} in the network---essentially its function---is a strong indicator of how and with what other systems it communicates.  We build role-based models of systems that allow us to identify connections that deviate from patterns suggested by role; such anomalous connections might indicate adversarial lateral movement. 

The first approach deals with the detection of generic lateral movement: both exploit-based and ostensibly authorized (authenticated and otherwise legal) but possibly malicious connections between pairs of systems on the network.  It is based on the observation that, while a particular system might make connections with a large number of unique systems over some historical period (say, several weeks), it makes connections with a smaller and more stable number of {\it types} of systems.  By ``type'' of system we have in mind the functional {\it role} of the system, so, for example, authentication servers (like Domain Controllers on a Windows network), mail servers, DNS servers, and workstations are prototypical system roles. We find that systems tend to communicate with the same system roles over time.  Meanwhile, an attacker will choose their movements based on efficiency or expediency, and might well link systems that have little business communicating otherwise.  Examples abound: IoT devices with broad connectivity and no access controls offer a stealthy route through the network; however, many systems like workstations and core servers might rarely (if ever) communicate with them. Attackers can establish pivots on any compromised system through which they can channel command and control, route commands, or move data: these pivots might link systems that seldom directly communicate, {\it e.g.} workstation-to-workstation, file server-to-printer, mail server-to-workstation, and so on.  This {\it role-based} lateral movement detection applies to individual systems: it uses unsupervised methods to learn the roles of its peer systems and flags connections to systems with novel or seldom-seen roles.  


The second approach concerns connections between systems with roles that {\it do} normally communicate, for example, workstations and authentication servers.  All inter-system connections are initiated and maintained by processes running on the hosts involved: we hypothesize that, in general, normal process dynamics follow patterns based on the roles of the systems involved in the connection.  Normal system functions, like authentications, NTP syncs, and DNS queries, tend to involve recognizable combinations of processes over time; for example, when a Windows client authenticates against a Domain Controller, the process \texttt{lsass.exe} initiates the connection and is often followed by processes like \texttt{svchost.exe} or \texttt{ntoskrnl.exe} as dynamic link libraries and drivers are loaded into memory to facilitate post-authentication tasks.  Meanwhile, Domain Controllers interact with NTP servers or DNS servers for entirely different purposes, and so involve different processes and/or process dynamics.  

Such patterns of normal behavior can be useful for spotting malicious inter-system activity, particularly that involving process injection and hijacking, which might be evident as deviations from these patterns.  For example, an attacker might wish to use custom malware to establish a remote connection but opts to hide this code by injecting it into an already running executable, like \texttt{chrome.exe}.  That seems innocent enough, but this action might disrupt this executable's normal dynamics (by causing it execute along with processes it doesn't typically associate with when connecting to systems of the given role);  alternatively, the process could be one that seldom supports connections to systems of a given role (for example, \texttt{powershell.exe} opening a connection between two workstations).  

This {\it role-process-based} detection also applies to individual systems, where a different model of process dynamics is learned for each of its peers' roles.  Each model uses compression based on frequent-itemsets to recognize common process patterns expected of the role in question, and identifies rare patterns as indicating potentially malicious connections.


In summary, we propose an unsupervised framework that seeks to identify lateral movement on two levels.  It essentially asks: should these two systems be talking?  If so, are they talking in the correct way?   
The notion of system {\it role} is central to this combined framework, and should be appreciated as supplanting the individual system as the entity that is being profiled.  
\section{Related Work}
This analysis concerns the detection of malicious connections within enterprise networks.  Though generic, in the sense that the method is agnostic to protocol, technique, or payload, its benefit is most clearly seen as a means of detecting authenticated connections, for which explicit indicators of compromise and other rules are generally lacking.  It is therefore applicable to the authenticated lateral movement by an external adversary via compromised accounts, but also to malicious insider and masquerade activity, which likewise proceed via authorized means.  

There are two broad areas of prior research relevant to this work.  The first has to do with system classification, by which we group systems into roles, and is central to our {\it role-based} detection models.  The second has to do with the analysis of sequences of system events (user activities, system calls, process executions) for anomaly detection, relevant to our {\it role-process-based} detection models.  We review the related literature for each separately below.

\subsection{System classification}
The unsupervised organization of systems into classes, or types, is central to both of the detection methods outlined in this paper.  There are many ways to perform this categorization: we opt for a grouping based on system role---how the system behaves in terms of services provided to other hosts on the network.  This method makes use of only a single source of passive connection log data, and employs standard clustering algorithms on features derived from these data.  It is therefore a simple {\it example} of how the central concept, one emphasizing the importance of system type in detecting malicious connections, might be implemented in an operational setting.  We therefore don't present our system classification approach (described in the next section) as novel: indeed, many of the elements it incorporates have been studied previously.  The references that follow review the basic literature and suggest alternative conceptions of system type that could be implemented in our framework.   

The problem of system and traffic identification, in terms of application, protocol, or service type, has seen widespread development, primarily for applications to quality of service, policy enforcement, and network management.  We emphasize that we are not looking to perform traffic classification, that is, the assignment of traffic flows to known protocol or application categories like peer-to-peer communications, online gaming, or TOR \cite{NguyenT,KimHy,Salman}.  Nor are we interested in performing traffic anomaly detection (to identify things like worm propagation or port scanning) using traffic classification (see, for example \cite{Lakhina,McHugh} and references therein).  Our problem is considerably easier, in that we do not care to positively identify the ground truth label of a particular group of systems; we simply wish to group them together according to functional characteristics.  In this light, we focus on previous literature with similar goals.

The use of network flow data formed the basis of several early studies.  The important work of \cite{XuK} performed clustering on the (\texttt{src\_ip}, \texttt{src\_port}, \texttt{dst\_ip}, \texttt{dst\_port}) quadruplet to resolve systems into broad categories, like servers, proxies, and traffic related to scanning or exploits. Also working with flow records, \cite{Wei} analyzed traffic statistics like daily byte totals, number of distint destinations, and average TTL to cluster systems into broad categories like TCP servers, UDP-only servers, and user workstations.  The related work of \cite{Erman} applied a variety of clustering algorithms to flow-derived characteristics like number of packets, mean packet size, and mean inter-arrival time of packets, to organize traffic according to common services like web, email, and peer-to-peer. The technique of {\it graphlets}, small undirected graphs yielding topological representations of host-to-host communications using IP and port data, was used for supervised traffic classification in \cite{KaragiannisBLINC}.  Graphlets have been applied to the problem of unsupervised classification in \cite{Himura}.  

The unsupervised resolution of enterprise systems into roles based on connection patterns was studied in \cite{Tan}.  Hosts that are similar with respect to their neighboring system sets are grouped together hierarchically; the success of this approach hinges on the extent to which functionally similar systems are also similar with respect to this metric.  In \cite{XuKuai}, a graph of system associations is introduced where system vertices share an edge if they have made connections to the same remote system; edges are weighted according to the cardinality of this shared neighbor set.  Spectral clustering is then performed on this graph to organize systems into groups with similar neighbor interactions.  Dewaele {\it et al.} \cite{Dewaele} perform clustering on nine connection-related quantities, including things like number of peers, ratio of the number of destination ports to number of peers, and ratio of the entropies of second and fourth bytes of destination IP addresses.  This approach successfully categorizes traffic by common protocols, like web, peer-to-peer, mail, and DNS.   

Rather than perform role analysis by proxy (via connection topologies and other characteristics), we use connection log data to instead categorize systems by port usage: this kind of profiling gets closer to a functional description than connectivity patterns.  But, as stated earlier, there is no {\it a priori} reason that the kinds of categorization emphasized in the above works cannot be used in this approach.  

\subsection{Lateral movement and masquerade detection}
There is a tremendous body of research on the problems of lateral movement and insider threat detection, summarized in the following reviews \cite{Salem2008,Bertacchini09asurvey,LiuLiu,Bridges}.  Much prior research in this area has focused on the detection of exploitation and malware activity \cite{Bhuyan,Drasar,Tavallaee}, generally in the context of the KDD CUP 99 intrusion dataset.  In contrast to that diverse body of work, the method we develop here is a behavior-based detection capability aimed at generic lateral movement with no explicit indicators of compromise.  
\subsubsection{Network-based indicators}
A relevant body of research focuses on malicious authenticated accesses (or login events) inside the network, applicable to both insider and outsider threats.  Authentication graphs, which represent systems as vertices and logins as edges between them, were introduced as a tool for detecting anomalous login activity in \cite{Kent}.  These graphs can be constructed from authentication log data, and have been analyzed in a number of works in both supervised \cite{Kaiafas,Bai,Goodman} and unsupervised settings \cite{Holt,Siadati,Chen,Eberle,Eberle2,Powell,Bowman}.  Some authors have augmented login data with other data sources to characterize lateral movement: \cite{Chen18} combine authentication records with data on general network connections and DNS queries to create graph features that are used to train an autoencoder to perform anomaly detection. In \cite{Bian} data from network flows are combined with authentication records to build a supervised classifier to detect malicious connections.  Drawing connection and command \& control data from widely deployed monitors across the network, \cite{Bohara,Fawaz} create graphs of all network connections and seek to identify long chains of connections indicating multi-pivot lateral movement.  Long chains of pivoting activity are also the target of the flow-based detection schemes presented in \cite{Apruzzese,Husak}.   Bipartite user-system graphs created from multiple data sources, including login, web access, email, and file access records, are used to train a one-class learner in \cite{Gamachchi1,Gamachchi2} to identify malicious connections.  In \cite{Yen}, data from multiple sources, including authentication and proxy logs, and connection-oriented data like user agent strings, are used as features that are clustered to identify anomalies; this approach applies to both malicious external and internal connections. Application- and technique-specific methods have also been explored: in \cite{Djidjev}, graphs representing secure shell (SSH) connections are mined to identify subgraphs corresponding to single user activity, where large subgraphs of low probability are flagged as anomalous.  In \cite{Purvine}, reachability graphs representing logical routes through the network are analyzed to identify at-risk systems as those with high importance and high reachability; these insights can be useful for devising mitigation strategies.  

In contrast to these works, our framework assesses individual connections (versus chains of accesses) for anomaly,  where the connections can be of any kind (versus primarily authentications).  Importantly, though, our approach should not be viewed as an alternative to these methodologies, but instead as potentially mutually reinforcing since each targets related by different aspects of the lateral movement problem.

\subsubsection{Host-based indicators}
Another major body of work is focused on host-based indicators of malicious activity, and includes the analysis of system call traces, command line usage, file access patterns, and other user-driven behaviors.  Many of these studies make use of categorical and discrete sequence anomaly detection schemes that relate to, and contrast with, our method of identifying anomalous process clusters in time series.  The following references are most relevant to our role-process-based approach to lateral movement detection (see also the recent reviews \cite{Bridges,LiuM}).

The earliest works in this area applied association rule learning to sequences of Unix commands to create models of user behavior \cite{Teng,Forrest,Lee}.  The basic idea is that a user's shell commands follow a pattern, and that deviations from this pattern might indicate that the account has been compromised, or that the user is a malicious actor. The latter two studies explored the use of RIPPER \cite{Cohen} to identify anomalous commands not predicted by the rules.  Unix command analysis would go on to serve as the basis for a great number of further studies.  In \cite{Davison} a method, named Ideal Online Learning Algorithm, assumes that a Markov process governs command sequences and applies exponential weighting of past Unix command data to predict the next command; anomalies can be identified as prediction errors. Naive Bayes is used in \cite{Maxion} to ascertain which user of a closed set generated a certain command sequence; it is effective at this problem but it is not apparently designed to handle novel users (like an outsider threat).  Hidden Markov Models (HMM) were soon applied to the problem of modeling command sequences: in \cite{Lane}, HMM was applied to Unix commands and compared against Naive Bayes, revealing only a slight preference for HMM.  This suggested that indeed there might be important temporal structure in command sequences that can be modeled probabilistically; however, the analysis \cite{Yeung} concluded that HMMs were inferior to simpler frequency-based analyses.  Further, \cite{Iglesias2} argues that $\chi^2$ testing of command subsequences is superior to HMMs unless the subsequences contain $\mathcal{O}(1000)$ commands.  Similarity comparisons of fixed-length command sequences were explored in \cite{Lane2}, where it was found that profiled users can be accurately differentiated by their command behaviors.  A combination of Naive Bayes and similarity measures was proposed in \cite{Sharma}, where similarity based on Gaussian kernels was shown to yield general improvement over Naive Bayes.  Cosine similarity was explored in \cite{Iglesias}, where fixed-length sequences of user commands are compared against a prototype sequence from each class of user.  A comparison study of several different probabilistic models is found in \cite{Schonlau}, where it is reported that hybrid multi-step and Bayes one-step Markov processes are the best models of Unix command sequences.  In \cite{Balajinath}, genetic algorithms are employed to model fixed-length user command sequences, with anomaly scores influenced by the proportion of correctly predicted commands in the sequence.  

The other major subject of behavioral profiling useful for the detection of masquerades and account compromise is the system call trace: the sequence of processes used by a program to interact with the system's kernel. Malicious activity is expected to alter these sequences, and so models trained on normal system call sequences can be useful for anomaly detection.  Just as program execution gives rise to a sequence of system calls, so too do network connections give rise to a sequence of inter-system processes.  As we study the latter in this paper, several of the following approaches were consulted and will be discussed in connection with our use-case throughout the paper.  

The earliest studies in this area are perhaps the works of Hofmeyr and collaborators, \cite{Kosoresow,Hofmeyr}.  In these works, sequences of system calls are divided up into $k$-length subsequences and compared via Hamming distance, where this distance serves as an anomaly measure.  In subsequent works, this technique is referred to as {\it sequence time delay embedding} (stide); we will have more to say about this approach later in the paper.  In \cite{Warrender}, stide was improved by replacing the Hamming distance with a simple mismatch count across the subsequence and compared with RIPPER and HMMs, where it was found that HMMs enjoyed the lowest false positive rate.  Soon, recurrent neural networks \cite{Ghosh} and evolutionary neural networks \cite{HanSJ} were brought to bear on the problem: both neural networks outperform stide. Neural networks with radial basis function units were considered in \cite{Ahmed}, where particular attention was paid to the window size defining individual attack subsequences among the longer sequences of system calls.  The window size in this work corresponds to $k$ in the stide analyses, and, as with stide, it is found to have a strong influence on accuracy. The fixed-length comparison window was done away with in \cite{Eskin}, where sparse Markov transducers were used to identify context-dependent window sizes; these models generally outperformed stide and related fixed-window size methods.  Further work to identify meaningful groupings of system calls was conducted in \cite{Creech}, where an extreme learning machine was able to learn to group calls into appropriate semantic units; this model outperforms stide and HMMs.   

An approach that builds dictionaries of anomalous system call sequences (rather than working to model normal sequences) is developed in \cite{Cabrera} using stide to identify them; this approach performs best in the context of specific Unix programs, like {\it sendmail}, where anomalies are circumscribed by the process.  In \cite{Xie}, a one-class support vector machine is trained on a set of labeled malicious system call data, including common exploitation tools like the Metasploit meterpreter and hydra login cracker.  It performs well, but is not an anomaly detection system.  Support vector machines based on sequence-similarity kernels are shown to outperform radial basis function kernels in \cite{TianS}.  If sequences are interpreted as documents, they can be analyzed using method of text categorization: in \cite{Liao}, individual system calls are {\it tf-idf}-weighted and assessed for anomaly with $k$-nearest neighbors distance; in \cite{Rawat} system calls are binary weighted with comparable results to \cite{Liao}. Following in this vein, \cite{Chen2005} apply support vector machines and neural networks to these features in a supervised setting.  A ``bag-of-system calls'' representation is adopted in \cite{Kang}, which reports superior performance over fixed-length method like stide.   Similar frequency-oriented system call representation were analyzed in \cite{Ye} with a variety of statistical techniques, like $\chi^{2}$, Hotelling's $T{^2}$, and Markov chain-based tests.  

Rule learning is applied to system call traces and compared against stide in \cite{Tandon}, where it is shown to do better than stide at detecting true positives labeled according the 1999 DARPA attack taxonomy. An analysis based on frequent itemsets is developed in \cite{Chen2006} and shown to outperform a support vector machine trained on system calls of different users.   Hidden Markov Models are explored in \cite{QianQ} who find the method ``practicable'' but note that proliferation of hyperparameters makes these models difficult to tune; various speed-ups were explored for HMMs in \cite{HoangXD,HuJ} resulting in efficient and accurate models.  A probabilistic model based on kernel states is developed in \cite{Murtaza}, and shown to yield lower false positive rates than stide and HMMs.  An interesting model is developed in \cite{Tapiador} that uses information theoretic measures to identify attackers that are trying to deceive anomaly detection capabilities, through such actions as command padding.  In \cite{Maggi}, analysis of system calls together with their arguments, initiated in \cite{Kruegel,Tandon}, is expanded by first clustering system calls into classes, and then using these classes to build an HMM model to recognize anomalous calls.  This approach is conceptually similar to ours, which also employs clustering to group together processes occurring as part of larger meaningful functions; in our case, though, we seek temporal clusters whereas in \cite{Maggi} the intent is to group together calls with similar arguments.   

Other aspects of user behavior have been modeled for anomaly detection: \cite{Salem} model user actions related to file and information access, with the expectation that adversaries (making use of compromised accounts) won't be as directed and efficient in this task.  Other characteristics of file system access, like timestamps and file size, were analyzed in \cite{Mehnaz}.  GUI interactions, including keyboard activity and mouse movements were modeled via SVM in \cite{Garg} and random forest applied to Microsoft Word interactions in \cite{Masri}.  Recurrent and convolutional neural networks were employed in \cite{Singh} to model the temporal behavior of various user behaviors, like logon times, and types of applications and amounts of data accessed, to detect anomalies.  These models perform comparably the collection of methods discussed in \cite{Lazarevic}.

We now briefly reflect on some of the methodologies above in the context of role-process-based lateral movement detection.  Methods that rely on fixed-size subsequences, or that don't tolerate subsequences of arbitrary sizes, will not perform well against our use-case.  Fixed-sized subsequences include processes that can be arbitrarily far apart in time, and hence causally unrelated.  Such subseqeucnes won't correspond to higher-level system functions and won't exhibit the associated regularity.  We verify that stide, with its fixed window size, performs poorly against our use-case.  Additionally, association rule learning fails for the simple reason that there isn't a reliable way to handle subsequences of only a single element, a common situation for our problem (corresponding to single isolated processes).  Finally, HMM and recurrent neural networks are high-quality temporal models; however, in this case we only wish to understand the temporal structure of individual process clusters, since these correspond to the higher-level system functions that might be disturbed by the adversary.  These process clusters are small time series, with the majority containing fewer than five elements, far too few to reliably train these kinds of models or to contain the rich temporal dependencies worthy of their power. 

\section{Role-based anomalies}
We begin with the premise that a particular system on an enterprise network generally makes use of the same networked resources over time: for example, a workstation will authenticate to the domain, and its user will send some emails, browse a file share, and complete a time sheet on an internal web server.  The remote systems to which it connects to perform these tasks---its {\it peers}---each assume one of a relatively stable set of {\it roles}.  In this example, the workstation regularly accesses a file share, say, but it never makes connections to email gateways, other workstations, or virtualization servers.   The adversary, meanwhile, might make connections to systems with {\it any} role, so long as the move supports their objectives.  Role-based anomaly detection applies to an individual system (hereafter referred to as the {\it subject} system) and learns the roles of its peers over time.  When a connection is made between the subject and a peer system with a novel (or perhaps seldom-seen) role, it is flagged as anomalous.   

This method proceeds under the assumption that the ground truth role of each peer system is unknown, and must instead be inferred from connection data.  This avoids the need to manually assign each system on the network to one of a pre-defined and comprehensive set of roles.  To organize a subject's peer systems---its {\it neighborhood}---into roles, we propose a system classification method based on port usage: each system is given a {\it server profile} based on the local ports it uses to {\it serve} data to other systems over some time period. Systems can additionally be given a {\it client profile} based on the remote ports of servers accessed by the system over some time period.  These two profiles are meant to summarize essentially how a particular system acts as a server and how it acts as a client.  These profiles take the form of vectors in ``port space'' with values quantifying the usage of the port.  The term {\it usage} is fairly general: one could consider the relative amount of data sent over the port, the relative number of connections involving the port, or something else.  A more coarse-grained representation would be a simple binary ``on/off'' for each port.   

Roles are assigned by clustering the server and/or client profiles of peers in the subject's neighborhood: systems whose profiles belong to the same cluster are considered to have the same role.  Roles are never positively identified (that is, given descriptive labels like ``Domain Controller''); they are merely bins for organizing systems.  Furthermore, a peer system's role depends on the other peers in the subject's neighborhood (since the number and quality of clusters depends on the dataset); as a result, the same system might assume two different roles in the neighborhoods of two different subjects.  

One important caveat of this approach is that, while servers of a certain type (say, Domain Controllers) need not make use of ``standard'' ports for its functions (because we are not interested in positively identifying the system role), they all need to make use of the {\it same} ports for the same functions ({\it e.g.} Kerberos must be listening on the same port on all Domain Controllers, whatever it happens to be).

We discuss this procedure in more detail below.   

\subsection{Creating a server profile}
To create a server profile, a data source is needed that records information about port-to-port connections involving the subject system.  Relevant data fields are the IPs of the systems and the ports involved in the connection, and timestamp.  Network traffic data, like Netflow, and endpoint logs, like Carbon Black Network Connect (netconn) data, are two suitable sources that provide these data.    In this study, we make use of netconn data collected on a large number of endpoints within a real, operational enterprise network; specific fields of interest are \texttt{local\_ip}, \texttt{remote\_ip}, \texttt{local\_port}, \texttt{remote\_port}, and \texttt{timestamp}.

In what follows we walk through how to build a server profile for a single system, which we take to be a Domain Controller (DC) on a Windows network.  First, the  netconn database is queried for all records with either \texttt{local\_ip} or \texttt{remote\_ip} matching the DC's IP address.  We are interested in records across a time period sufficiently-long to capture a representative sample of connections; time enough to include several thousand records is a good rule of thumb, but some experimentation might be needed.  

\begin{figure}[ht]
\centering
\includegraphics[width=3.5in]{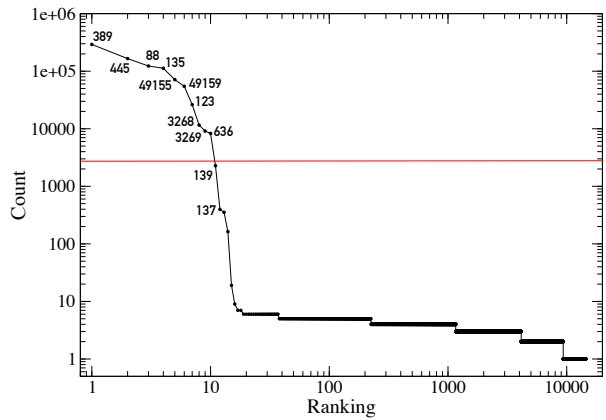}
\caption{\footnotesize{Domain Controller ports ranked according to the number of connections they were involved in over a 24-hour period. Only the most-used ports are numbered.  Only those ports above the red line will be used to profile the system.}}
\label{DC_server_ports}
\end{figure}

\begin{figure}[h]
\centering
\includegraphics[width=3.0in]{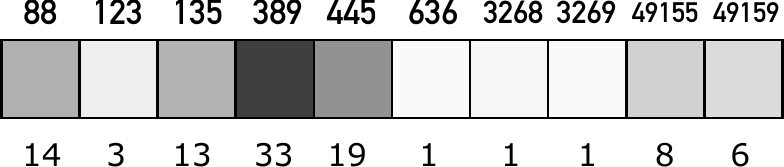}
\caption{\footnotesize{Domain Controller server profile schematic.  Port numbers across the top label the boxes, which are colored according to the percentage of all connections they were involved in over a 24-hour period.  Percentages are provided numerically below each box.}}
\label{DC_ports_schem}
\end{figure}

To create the DC's server profile, we must identify which ports local to the DC are serving data to peer systems.  We start with the list of all DC ports found in the record: these are the \texttt{local\_port}s when the DC IP matches \texttt{local\_ip} and the \texttt{remote\_port}s when it matches \texttt{remote\_ip}.  This list obviously includes both client ports (those ports local to the DC that are initiating connections to remote systems) and server ports (those ports local to the DC that are receiving connections initiated by remote systems).  Client ports are generally ephemeral, high-numbered ports that are chosen at random when the DC initiates a connection; as such, we do not expect to see many instances of such ports in the record.  We therefore implement a basic heuristic for deciding which ports are most likely server ports: we rank ports in descending order by number of connections in the record.  Figure \ref{DC_server_ports} shows this ranking for the DC. 
Notice that there are only appreciable numbers of connections associated with the first dozen or so ports: these are almost certainly server ports.  We impose a simple rule to collect only the most-used ports: we keep a port if its connection count is within a fixed factor of the count of the consecutively higher-ranked port.  In this study, we select a factor of three, which corresponds to the red horizontal line in Figure \ref{DC_server_ports}.  For example, port 636 had 8289 connections giving a cut-off of 8289/3 = 2763 for the next highest-usage port; since port 139 had only 2282 connections, it is not included.  With this rule the system's server profile is based only on the highest-usage server ports, and client ports are eliminated.  Figure \ref{DC_ports_schem} gives a schematic representation of the DC's server profile in terms of connection proportion: we can readily verify that these are standard ports associated with the major server functions of the DC ({\it e.g.} Kerberos authentication, network time protocol, lightweight directory access protocol, remote procedure calls, etc).
The size of the DC's server profile is typical of other servers---at most a dozen or so dimensions, making it a succinct summary of how the system acts as a server. 
\subsection{Creating a client profile}
The creation of a system's client profile parallels the above discussion, but here we are interested in understanding how the system tends to act as a {\it client}.  For this purpose, we focus on the remote ports accessed by the system, which are the \texttt{local\_port}s when the DC IP matches \texttt{remote\_ip}, and the \texttt{remote\_port}s when it matches \texttt{local\_ip}.  We again need to eliminate ephemeral ports on the remote systems (which are associated with {\it their} client activity) so that only remote server ports are considered.  The same usage ranking heuristic employed to create the server profile is useful here as well, but we won't repeat the details. The end of this process results in a client profile vector akin to the server profile vector of Figure \ref{DC_ports_schem}. 

\subsection{Identifying system roles}
\begin{figure}[h]
\centering
\includegraphics[width=3.0in]{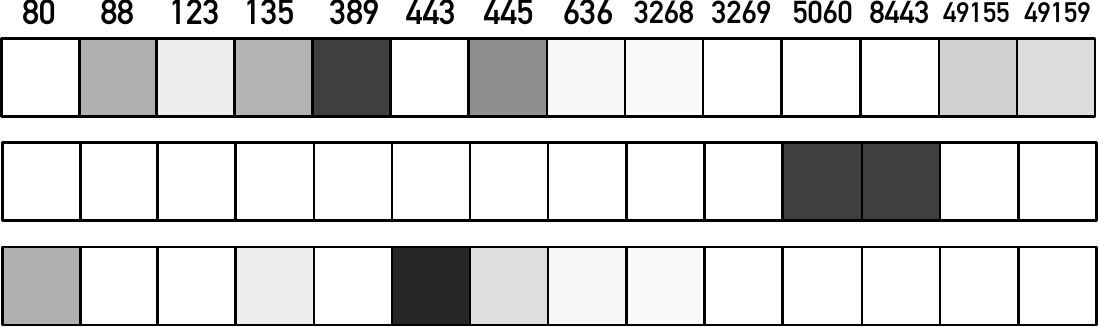}
\caption{\footnotesize{A group of server profiles.}}
\label{group_ports_schem}
\end{figure}

To organize a subject's peer systems into roles, we first must identify all of its peers over some time period.    Since we are focused on detecting lateral movement, which is internal to the system's network, only systems internal to the network are considered.   Once the list of peers is obtained, each one must be profiled using the above method.  Then, similar profiles must be grouped together: these groups are the system {\it roles}.  In Figure \ref{group_ports_schem}, a small example neighborhood of three peer systems is shown: the DC from above, a virtualization server (with active ports 80, 135, 443, and 445), and a VoIP server (with active ports 5060 and 8443) in the community.  Notice how the DC's profile has been expanded to include (zero) entries for the other systems' ports (and likewise for each of them); this is done so that the peers can be clustered together. 
We explore several clustering techniques and profile representations to find those that work best for this application.  
Two profile representations are considered: the one discussed in section 3.1 and depicted in Figures \ref{DC_ports_schem} and \ref{group_ports_schem}, in which each port is assigned a value between 0 and 100 corresponding to the percentage of connections involving the port over the course of the historical record.  The other representation is the more coarse-grained binary: a simple ``open'' or ``closed'' value for each port.  In what follows, the former is referred to as the {\it proportioned} and the latter the {\it binary} representation.  

\subsection{Partitional clustering of profiles}
The approach to system classification pursued in this analysis is independent of any particular clustering scheme, and so a variety of methods are explored using the two feature representations introduced above.  The first method considered is partitional clustering applied to the profiles, where each profile is a vector in a space with a dimension given by the number of ports represented across the neighborhood; in the example of Figure \ref{group_ports_schem} the dimension of the feature space is 14.  In practice, this space can become relatively high-dimensional, with generally several dozen ports represented; for example, the workstations in our data sample have an average 50 or so ports represented across their neighborhoods.  As is well known, distance-oriented algorithms can perform poorly and counter-intuitively in high dimensional spaces.   Care must therefore be taken when applying partitional clustering algorithms based on Euclidean distance, $d(x,y) = ||x-y||^2$, like $k$-means or mean-shift clustering.  To ameliorate the effect of high-dimensionality on clustering, there is a {\it spherical} $k$-means algorithm \cite{Hornik} in which the closeness of vectors is instead measured via the {\it cosine similarity},
\begin{equation}
d_\theta({\mathbf x},{\mathbf y}) = \cos \theta = \frac{{\mathbf x}\cdot{\mathbf y}}{||{\mathbf x}||\,||{\mathbf y}||}.
\end{equation}
The spherical $k$-means algorithm normalizes all vectors so that they are projected to points on the surface of a unit hypersphere.  These points are then partitioned on the surface of the sphere according to cosine similarity; this approach, in which points are considered close if they lie along similar directions from the sphere's center, has better high-dimensional performance than ordinary $k$-means using Euclidean distance as a similarity measure.  We test both $k$-means and spherical $k$-means in this study.
\subsection{Clustering profile similarity}
Rather than clustering the profile vectors directly, we can instead analyze the similarity matrix, ${\mathbf S} \in \mathbb{R}^{n \times n}$, consisting of pairwise similarities computed among all profile vectors in a neighborhood of $n$ systems. The matrix is normalized so that similar vectors $x,y$ have $S_{x,y} \approx 1$.  For proportioned data, cosine similarity is selected to give good high-dimensional behavior; for binary data, we in addition consider the Jaccard index applied to the systems' port sets (the set of ports appearing in each system's profile), 
\begin{equation}
J(x,y) = \frac{|p_x \cap p_y|}{|p_x \cup p_y|},
\end{equation}
where $p_{x,y}$ is the set of ports in profile $x,y$.  Once the similarity matrix is in hand, a variety of clustering methods can be applied directly to it.  We consider spectral and agglomerative hierarchical clustering as two conceptually distinct approaches to this problem.

Spectral clustering interprets the symmetric similarity matrix as the adjacency matrix of an undirected graph, and computes the graph Laplacian, ${\mathbf L} = {\mathbf D} - {\mathbf S}$, where the components of the degree matrix, ${\mathbf D}$, are 
\begin{equation}
D_{ii} = \sum_{j=1}^n S_{ij}.
\end{equation}
Next, the matrix ${\mathbf V} \in \mathbb{R}^{n \times \ell}$ with columns the $\ell$ most-relevant eigenvectors of ${\mathbf L}$ is constructed, and $k$-means clustering is applied to the {\it rows} of this matrix.  The relevance of eigenvectors can be ascertained by looking for the ``elbow'' in the plot of the corresponding eigenvalues; alternatively, the number of eigenvectors to include can be determined empirically using a quality measure of the resulting clusters. In what follows, we adopt the latter approach. 

Rather than cluster on its spectrum, hierarchical clustering applies directly the similarity matrix.  Agglomerative clustering begins with each datum in its own cluster, and then successively merges them into larger clusters.  First, the most-similar samples are merged.  A new similarity matrix is then computed based on these merged points (clusters), where similarity scores are computed among the new clusters using one of a variety of rules: here, we employ {\it average linkage} in which the new similarity score between two clusters is computed as the average of the pairwise similarities of all constituent points.  Clusters are continually merged until either all clusters below some similarity threshold have been merged, or, as with other clustering methods, the number of clusters can be determined empirically using a measure of cluster quality.  As with spectral clustering, we adopt this latter approach.

\subsection{Comparison of clustering methods}
The measure of cluster quality employed in this analysis is the {\it silhouette score}, which is the average of the silhouette coefficients, $s(x)$,  over all points, $x$, where
\begin{equation}
s(x) = \frac{b(x)-a(x)}{\max\{a(x),b(x)\}}
\end{equation}
where, for $x$ in cluster, $C$, 
\begin{equation}
a(x) = \frac{1}{|C|-1} \sum_{y\in C,x\neq y} d(x,y),
\end{equation}
is the mean intra-cluster distance $d(x,y)$ between $x$ and all other points $y$ in the cluster, and 
\begin{equation}
b(x) = \min_{C'\neq C} \frac{1}{|C'|} \sum_{y\in C'} d(x,y),
\end{equation}
is the minimum mean inter-cluster distance between $x$ and all other points $y$ not in $C$.  The coefficient satisfies $-1 \leq s(x) \leq 1$, with coefficients close to one indicating ``tight'' clusters.  In practice, the number of clusters, $n_C$, to use in a particular method is determined by computing the silhouette score over a range of $n_C \in [1,n]$ and selecting $n_C$ with the largest score.  

\begin{figure*}[ht]
\centering
\includegraphics[width=7.0in]{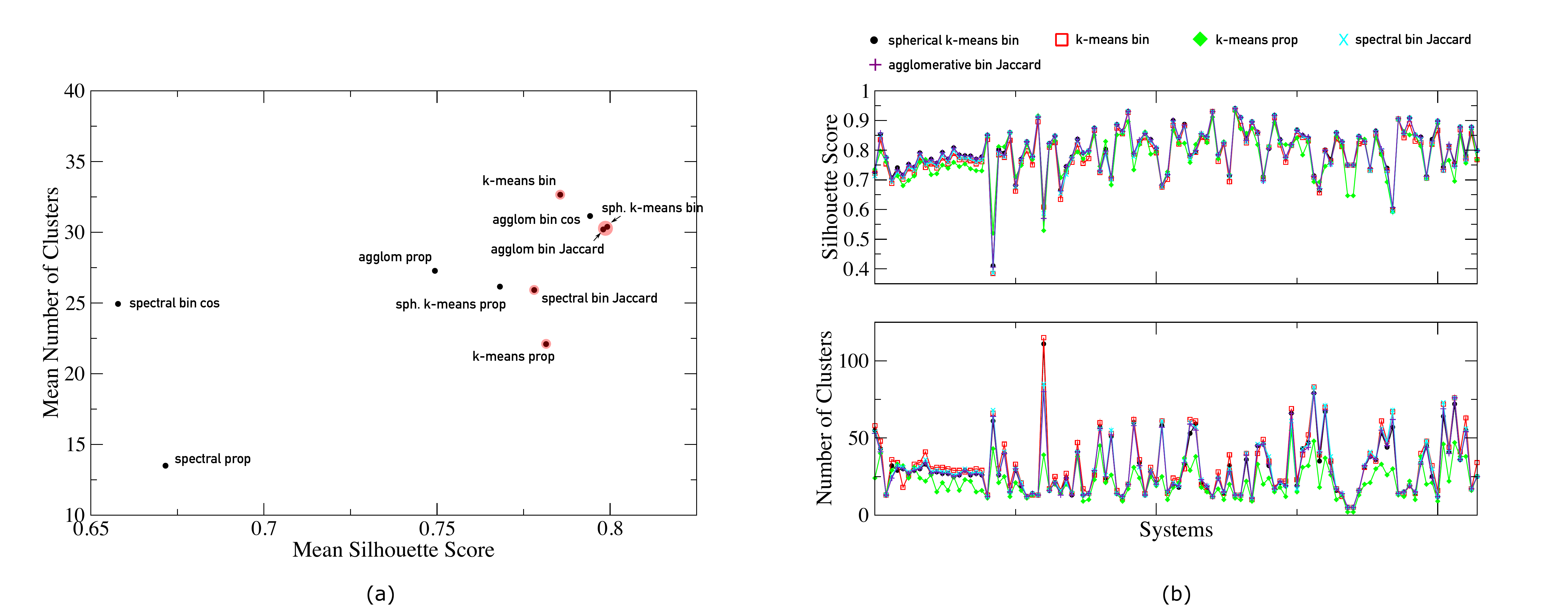}
\caption{\footnotesize{Results of applying multiple clustering approaches to the neighborhoods of 125 test subjects.  (a) Mean number of clusters and mean silhouette scores over all 125 subjects for each approach.  Those approaches highlighted in red are considered highest-quality (largest silhouette score).  See text for descriptions of the various approaches; abbreviations are as follows: ``prop'' refers to proportioned features, ``bin'' to binary features, ``cos'' to the use of cosine similarity in the similarity matrix, ``Jaccard'' to the use of the namesake index in the similarity matrix. (b) Silhouette scores and numbers of clusters found per test subject.}}
\label{clustering_res}
\end{figure*}
We take as our test data set a ``watch list'' of 125 high-value systems on the network that we would like to monitor for lateral movement.  These 125 systems include Exchange servers, Domain controllers, major application servers, and the workstations of highly-privileged users, existing on a real, operational enterprise network including thousands of hosts.  Each system on the watchlist is a subject: for each subject, the peers in its neighborhood are identified over a four week period via netconn records. We then test each clustering method on the proportioned and binary versions of these profiles.  For spectral and agglomerative clustering using the binary features, we construct similarity matrices using each of cosine similarity and Jaccard index; for the proportioned features, only cosine similarity is appropriate.  Each test system will have a different number of peers of different types, and consequently different numbers of clusters of varying quality.  

Results are presented in Figure \ref{clustering_res}: (a) shows the average number of clusters and the average silhouette scores over all test subjects for each profile representation (proportioned and binary) for each clustering method. In Figure \ref{clustering_res} (b), these quantities are broken out by test subject (listed along the $x$-axis) for the five methods highlighted in red in (a).  There are a few important things to note: first, with only a few exceptions (spectral clustering on binary features using cosine similarity and spectral clustering on proportioned features), all methods find fairly high-quality clusters (mean silhouette scores $> 0.75$).  Second, among these successful methods, we see good breadth in the average numbers of clusters, from around 22 for $k$-means on proportioned features up to 32 for $k$-means on binary features.  The fact that these methods all yield tight clusters indicates that there are natural clusterings on different scales ({\it i.e.} a cluster can be divided into two sub-clusters without hurting the silhouette score if that cluster stands in relation to other clusters the way its sub-clusters stand in relation to it.)  This is useful in practice because it allows one to vary the resolution of clustering, which affects the anomaly detection rate (we will look more closely at how shortly).  Also of note, as shown in Figure \ref{clustering_res} (b), is that the high-quality methods track each other well, suggesting that this featurization is quite robust to clustering method. 

\begin{figure*}[p]
\centering
\begin{sideways}
\includegraphics[width=9in]{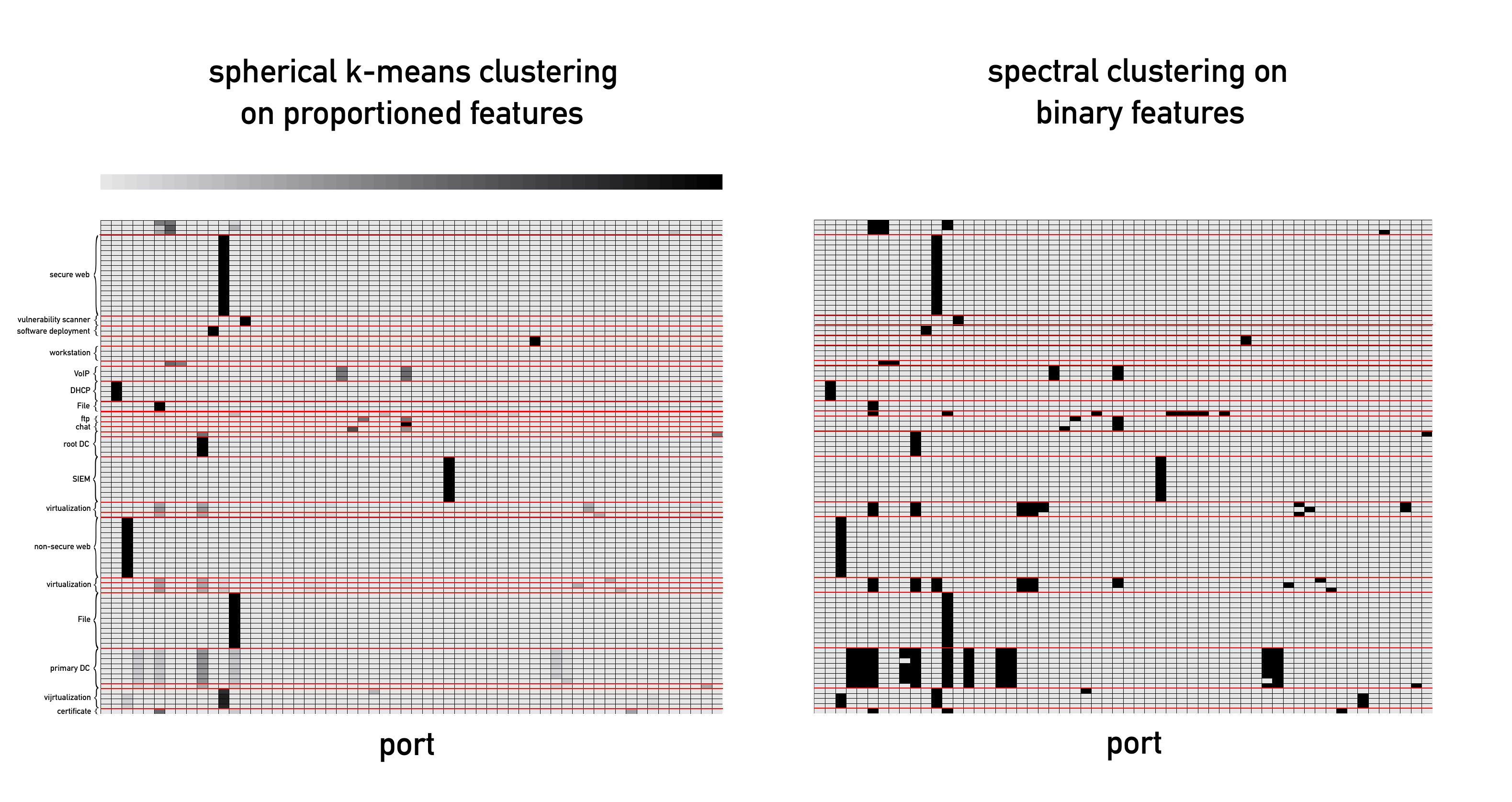}
\end{sideways}
\caption{\footnotesize{Detailed results of two clustering approaches: spherical $k$-means on proportioned features and spectral clustering on binary features using Jaccard index as similarity measure, for a single test subject.  Ports (unlabeled) run across the horizontal axis, peer systems run along the vertical axis.  Systems are grouped according to role, and shading of boxes indicates usage.  For binary features boxes are either black or white.  Red lines enclose clusters found by each method.}}
\label{port_clusters}
\end{figure*}

To examine the results of clustering more closely, in Figure \ref{port_clusters} we present the results of clustering the peers of one test subject (a workstation).  The clusters are superimposed on the peer server profile feature vectors for two different clustering approaches: (a) spherical $k$-means on proportioned features (with 27 clusters) and (b) spectral clustering on binary features (with 22 clusters).  Most systems in the figure have been labeled according to their primary function or server role as ascertained through discussions with administrators, use of standard service ports, or hostnames.  These labels serve as a putative ``ground truth'' against which to judge the clustering, though it is important to emphasize that not all systems with the same role will have the same behavior in terms of port usage.  For example, one root DC and one primary DC have notably different port profiles than the rest of their respective groups: spherical $k$-means clustering on the proportioned profiles separates these systems into their own classes, while spectral clustering on the binary profiles does not.  Which is preferable ultimately depends on how each effects the anomaly detection rate.  

Sometimes the finer clustering achieved with spherical $k$-means seems {\it too} fine: consider how the first two groups of virtualization servers are further broken up (the first group of three systems is resolved into two clusters, the second group of three systems into three clusters). Look closely at the second virtualization group: the first two open ports have close to equivalent usage across all systems in the group; the systems differ only in which high-numbered ports they have open.  Given that these ports have relatively low proportionate use, it is perhaps surprising that they are sufficiently important to affect the clustering.  In contrast, one might naively suppose that binary clustering would be {\it more} sensitive to single port differences among systems, since even low-proportion ports are given the full binary value of 1, exaggerating any differences.  But, examination of spectral clustering results on the binary features reveals this {\it not} to be the case.  Indeed, both groups of virtualization servers are given clusters corresponding to their roles.  While differences in open ports (however disproportionately those ports might be used) are emphasized, binary features also exaggerate the similarity of systems with open ports in common, regardless of any usage differences between them because all open ports are given equal values of 1. As long as systems have more open ports in common than not, clustering based on binary features will tend to group them together.  

\begin{figure}[h]
\centering
\includegraphics[width=3.5in]{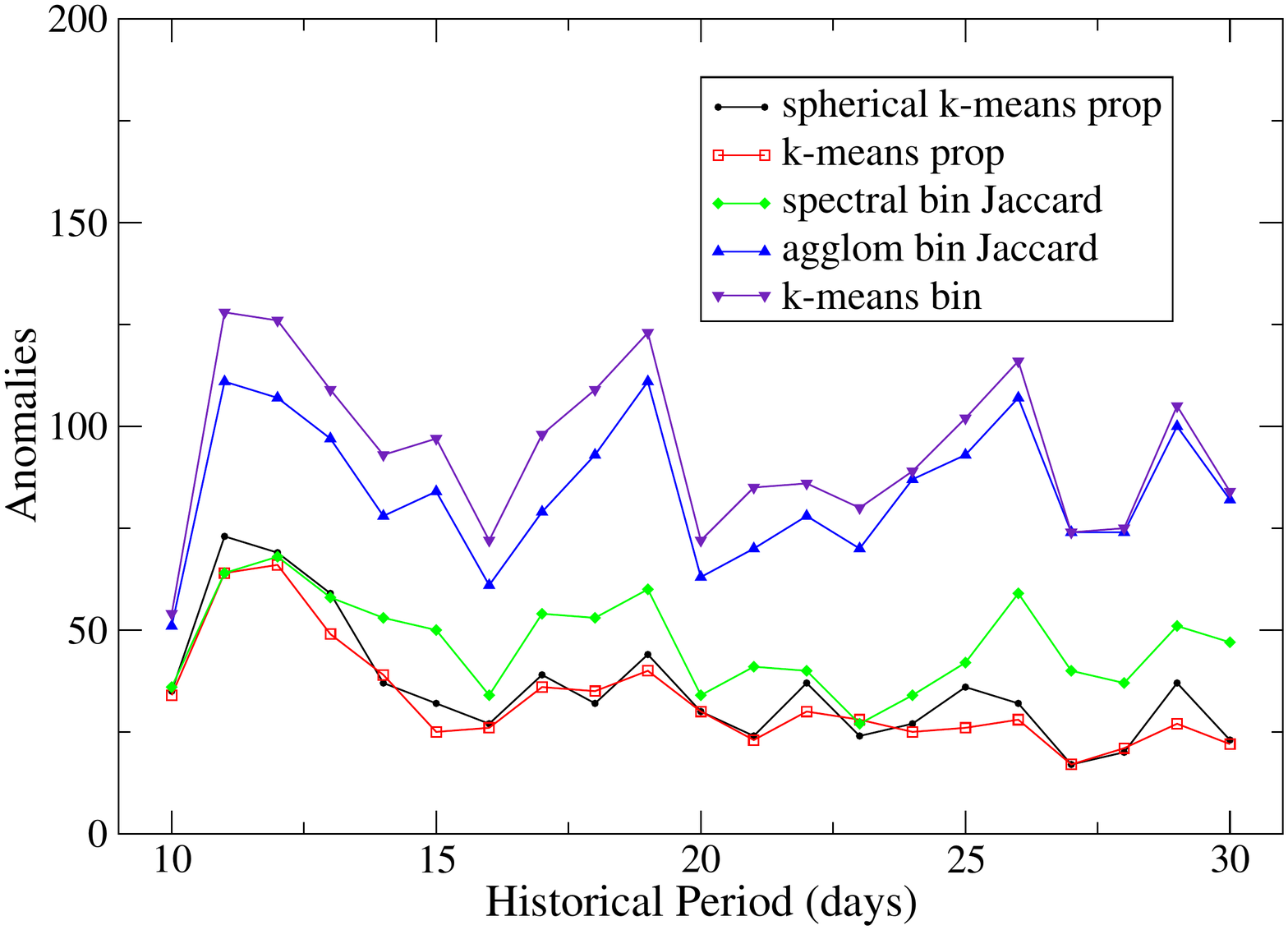}
\caption{\footnotesize{Number of anomalies found across 125 test systems as a function of length of the historical period for each of the five clustering approaches highlighted in red in Figure \ref{clustering_res}.}}
\label{clustering_hist}
\end{figure}

We now examine how the different clustering approaches might perform in daily operations on a real network.  We apply the 5 approaches highlighted in Figure \ref{clustering_res}: 1) spherical $k$-means on proportioned features, 2) $k$-means on proportioned features, 3) $k$-means on binary features, 4) agglomerative clustering on binary features with Jaccard index, and 5) spectral clustering on binary features with Jaccard index, to our 125 test subjects.  As before, we collected all netconn records of internal connections between these systems and their peers, but this time we vary the length of the historical record and study its effect on performance. We are interested in each subject's {\it new} peers: those systems whose only connections occur over the past 24 hours (this time period defining novelty is of course arbitrary).  Novel peer systems that occur in their own clusters, and hence have a novel role, are anomalies. There tend to be more roles in neighborhoods spanning longer time periods, and so we might expect the number of alerts per monitored system to drop as the duration of the historical is increased.  Figure \ref{clustering_hist} presents the number of anomalies identified (novel peer systems with novel roles over the most recent 24-hour period), aggregated over all 125 subjects, as the size of the historical record of each system is increased from 10 to 30 days.  

In keeping with earlier results, $k$-means on binary features, with its propensity for numerous clusters, results in the most novel clusters and hence most anomalies; it behaves similarly to agglomerative clustering on binary features using the Jaccard index as similarity measure.  Meanwhile, spherical $k$-means and $k$-means applied to the proportioned features perform comparatively, indicating that the dimensionality of the feature space is not so large as to benefit from the spherical $k$-means algorithm.  Ultimately, these results indicate that stability tends to be reached for each of the clustering methods after around 15 days of history.  With 30 alerts per day on average over all 125 subjects,  $k$-means clustering on proportioned features yields a manageable number of alerts, with $k$-means on binary features offering a more sensitive alternative.


\section{Role-process-based anomalies}
While role-based anomaly detection is able to spot unusual connections between the subject and systems with {\it novel} roles, it is not useful for identifying lateral movement between the subject and systems with {\it known} roles.  For this kind of connection, we introduce {\it role-process}-based anomaly detection.

Inter-system connections relevant to lateral movement are facilitated by {\it processes}.  By ``process'' we have in mind executable programs that are involved in maintaining connections and transferring data between systems.  The processes comprising standard communications tend to follow patterns, and we posit that these process patterns depend on the roles of the systems involved in the communication.  This is based on the premise that the role of a system determines what kinds of functions it requests or performs, and that these functions generally depend on the role of the other system involved in the connection.  For example, a workstation authenticates against a Domain Controller by opening a connection to port 389 (LDAP) via the process \texttt{lsass.exe}.  The Domain Controller in turn receives this connection with its own invocation of \texttt{lsass.exe}.  This authentication step might be followed by additional tasks; for example, if DLLs need to be loaded, the process \texttt{ntoskrnl.exe} will be invoked on the Domain Controller.  The combination \{\texttt{lsass.exe}, \texttt{lsass.exe}, \texttt{ntoskrnl.exe}\} in short succession might therefore correspond to a standard authentication operation. There are certainly variations on this theme: on networks with Windows Advanced Threat Analytics deployed, the Domain Controller immediately queries the authenticating client for threat analytics via the process \texttt{microsoft.} \texttt{tri.gateway.exe}, and so we might see this process sometimes included in the above sequence. Meanwhile, the interactions between the Domain Controller and an NTP server, or between two workstations, involve different processes or process patterns because the connections facilitate different functions. 

These patterns of normal operations, if they can be learned, can serve as a basis for discovering malicious activity. 
A popular method of lateral movement, and malicious activity more broadly, is process {\it injection} or {\it hijacking}, whereby the adversary runs an illicit process under the name and process identification (PID) number of a legitimate process, or enlists a legitimate process to execute or load illicit processes or libraries on the attacker's behalf.  These can be very subtle techniques, especially if the legitimate processes involved are very common and executed as part of a wide range of system functions.  If the legitimate processes involved in standard communications and system functions can be reliably profiled, then it becomes possible to potentially recognize illicit process injection or hijacking that alters these profiles.   For example, as we've seen \texttt{lsass.exe} frequently executes close together in time with \texttt{ntoskrnl.exe} during client authentication; if \texttt{lsass.exe} is instead used by the attacker for another purpose, we shouldn't expect to see the standard sequence of authentication-oriented processes execute on either system. Alternatively, lateral movement may progress via non-standard protocols (like psexec or dcom) or remote procedure calls invoked from customized \texttt{powershell} scripts following their own, possibly novel, patterns.  Against the backdrop of recognizable standard process patterns, these kinds of activities are expected to stand out. 

We therefore seek a means of mining patterns in process dynamics between a subject system and its peers in each role.  For each subject, we then have a process model for each role, {\it e.g.} Domain Controllers, file shares, web servers, and so on.  These models can then be used to test new connections to peers with known roles; for example, when the subject makes a new connection to a Domain Controller, we can compare the process sequence against those comprising the connections between the subject and all other known Domain Controllers over some historical time period.  
\begin{figure*}[ht]
\centering
\includegraphics[width=6.0in]{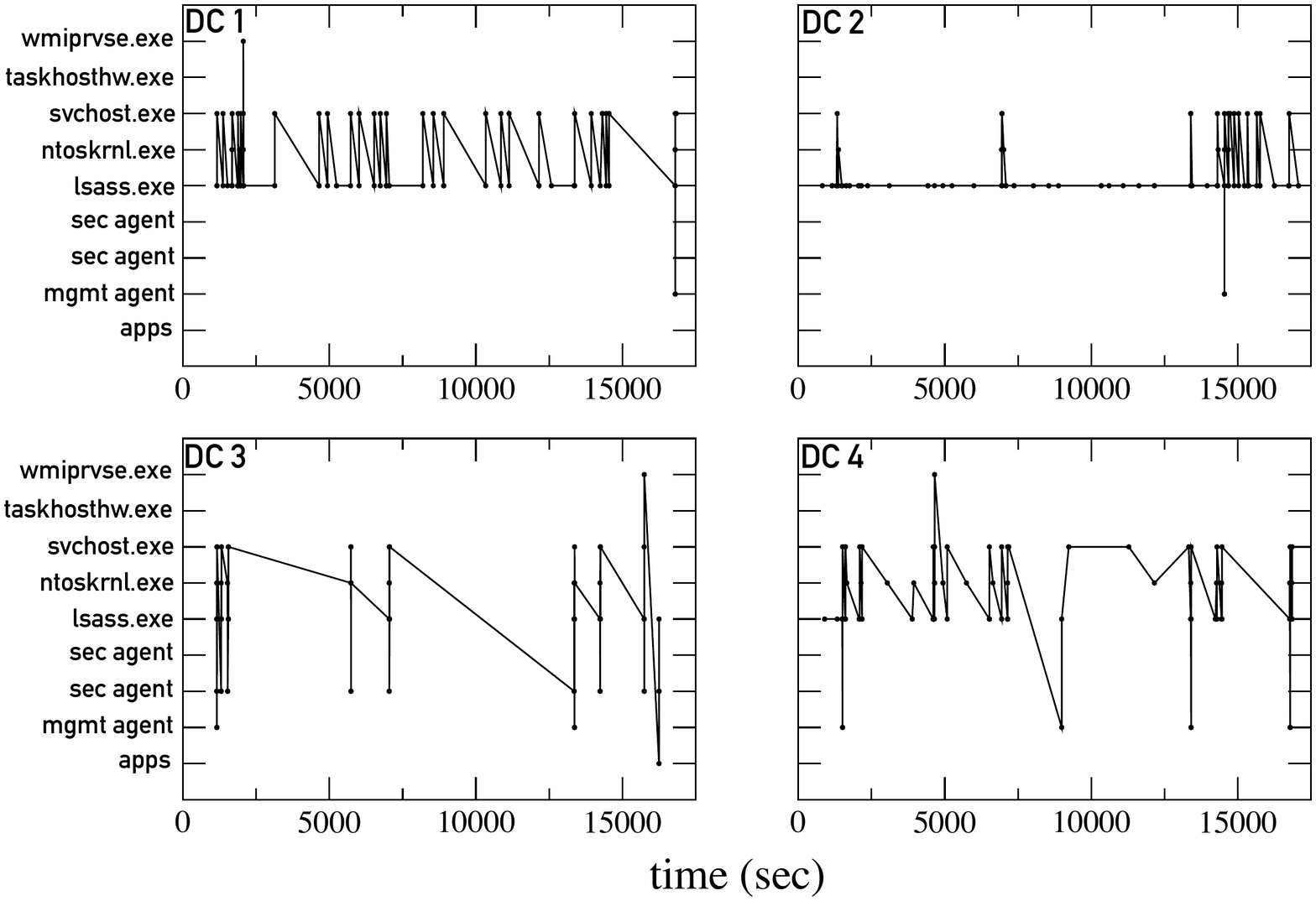}
\caption{\footnotesize{Time series of inter-system communications by process type between a single workstation and four Domain Controllers.}}
\label{DCs}
\end{figure*}

To make this tangible, Figure \ref{DCs} shows the time series of normal process activity between a workstation and four different Domain Controllers (DCs) over the same four hour period.   Rather than work with time series, the simplest model of process dynamics ignores temporal information and considers each process time series as an ordered sequence of processes.  This kind of modeling is similar to the analyses of Unix command line behaviors and system call traces discussed in Section 2, and so we apply the popular method of {\it sequence time delay embedding} (stide) used in many of those analyses to our process time series. Stide is a method of subsequence comparison: the test data is organized into subsequences of $k$ consecutive processes, and these subsequences are compared against a store of normal subsequences of length $k$.  Anomaly scores are therefore applied at the subsequence level: in \cite{Hofmeyr} the Hamming distance between test and normal subsequences was used, and in \cite{Warrender} {\it locality frame count} (LFC) was used.  The LFC is simply the number of mismatches between the test subsequence and the normal subsequences.  The choice of $k$ is arbitrary, though some authors suggest that $k=6$ is optimal in a wide range of applications \cite{1004371}. We choose $k=6$ in this analysis, but will see that model performance does not hinge on this value. 

To test a subject system with stide, we first need a collection (hereafter, {\it database}) of its historical connections to serve as our normal instances (this is the ``model'' for stide). The database is built from netconn data collected over some historical period: we fix this historical period at ten days.  For each peer system, the netconn data (which can be visualized in raw form as the time series of Figure \ref{DCs}) is translated into a sequence of processes.  In this study we wish to test all new connections arriving in a 24-hour period as a batch process (though these methods work just as well in streaming deployments).  The new records for each peer system are translated into process sequences, added to the peer's database, and then the whole database is segmented into $k$-length subsequences.   Next, all peer systems are assigned roles via one of the clustering methods detailed in the last section. Finally,  for each peer system, each $k$-length subsequence with new processes (called a {\it test subsequence}) is compared against all $k$-length subsequences in the historical databases of all peer systems with the same role.  The number of historical records that match the test subsequence serves as a simple anomaly score (with low scores indicating more anomalous subsequences).


To give stide a thorough investigation, we use it to test the same 125 high-value systems used to test role identification in the last section. We are first and foremost interested in the distribution of scores: we would like to understand how sensitive stide is to rare test subsequences, which tells us how well stide captures the normal process behavior of each role.  The $z$-score is computed for each subsequence, $z = (x-\mu)/\sigma$, where $x$ is the number of matches between the subsequence and those in the historical database with the same role, and $\mu$ and $\sigma$ are the mean and standard deviation of these records.  

\begin{figure}[h]
\centering
\includegraphics[width=3.5in]{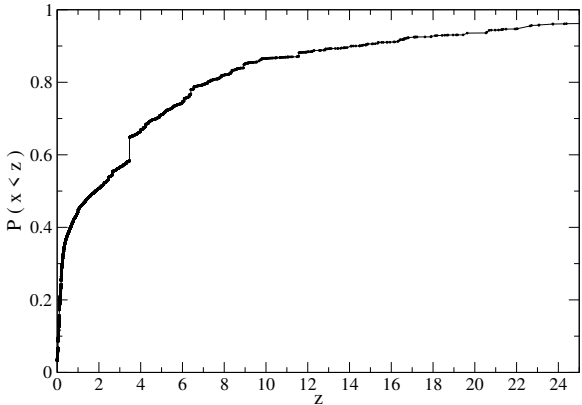}
\caption{\footnotesize{Cumulative probability distribution of test sample $z$-scores for stide, aggregated over all 125 test subjects for one 24-hour test period.}}
\label{stide}
\end{figure}

Figure \ref{stide} shows the cumulative distribution of the $z$-score, which gives the probability that a sample will satisfy $(x-\mu)/\sigma < z$.    
It is apparent that stide finds many subsequences with low numbers of matches: these are subsequences with large $z$-scores.  For example, around 15\% of the subsequences lie beyond $z = 10$; that is, a little over a tenth of subsequences are {\it very} rare (at least 10 standard deviations from the mean).  In practice, our anomaly threshold should be based on subsequence rarity, and so in order to keep the number of anomalies to a manageable level we are forced to consider only subsequences of the most extreme rarity.  For example, across our 125 test subjects, there are around 140,000 test subsequences in a 24-hour period: even with a rather extreme $z=10$ threshold on subsequence similarity, there will still be 21,000 anomalous records to investigate (around 150 per system).  This is too many to be useful in any realistic defensive scenario.  


One plausible reason that stide performs so poorly is that it discards all temporal information but process order:  adjacent processes in the sequence could be seconds or hours apart, and stide treats them identically.  It therefore mixes long- and short-timescale process dynamics, and there is unlikely to be much discernible order in such sequences.  The reasoning goes as follows: as a user interacts with a computer, discrete functions are performed: authentications, DNS queries, NTP syncs, file downloads, and so on.  For a given user, these events might occur roughly randomly in time (aside from possible regularities in such things as daily login times). When these functions are decomposed into their constituent processes, as we are doing here, this disorder persists and algorithms like stide struggle to identify any regularity. Recall the sample process subsequence corresponding to client authentication: \{\texttt{lsass.exe}, \texttt{lsass.exe}, \texttt{ntoskrnl.exe}\}.  With $k=6$, stide will never analyze this as its own subsequence, but always in combination with other potentially functionally-irrelevant processes.  
We therefore must find a way to first organize the longer process time series like those of Figure \ref{DCs} into more localized groups, or clusters, of processes (which are more likely to correspond to discrete system events and functions) and analyze {\it these groups} for anomalous behavior.  Sadly, stide was doomed from the outset: we explored it merely as a cautionary tale against treating process time series as flat, fixed-size sequences.

\subsection{Mining process clusters}
In this section we propose a way to isolate process subsequences that might correspond to higher-level system functions. To identify them in time series like Figure \ref{DCs}, we apply density-based clustering in the temporal domain to the process time series with a time threshold on the order of seconds.  The algorithm DBSCAN \cite{DBSCAN} creates dense clusters as follows:  1) a {\it core point}, which lies within a distance $\epsilon$ from at least {\it MinPts} other points (its $\epsilon$-neighbors) is placed in a cluster along with its $\epsilon$-neighbors; 2) if any of these $\epsilon$-neighbors is a core point, all of its $\epsilon$-neighbors are added to the cluster;  3) this process is repeated for all core points.   The choice of {\it MinPts} and $\epsilon$ are application-specific: since we are interested in process clusters of any size, we choose\footnote{Clusters of size 1 will still be found as noise by DBSCAN.} $MinPts = 2$. The appropriate choice of $\epsilon$ is less clear: one could base its value on some ``natural'' timescale of process dynamics, but this certainly depends on the function.   Instead, we elect to set $\epsilon$ equal to the typical distance between each point in the time series and its {\it MinPts} nearest-neighbors\footnote{In practice, this distance is plotted as an increasing function of points and the ``elbow'' of the curve is chosen for $\epsilon$.}.  In this way, $\epsilon$ is the natural distance under the assumption that the smallest clusters should have a size of {\it MinPts}.  

\begin{figure}[t]
\centering
\includegraphics[width=3.5in]{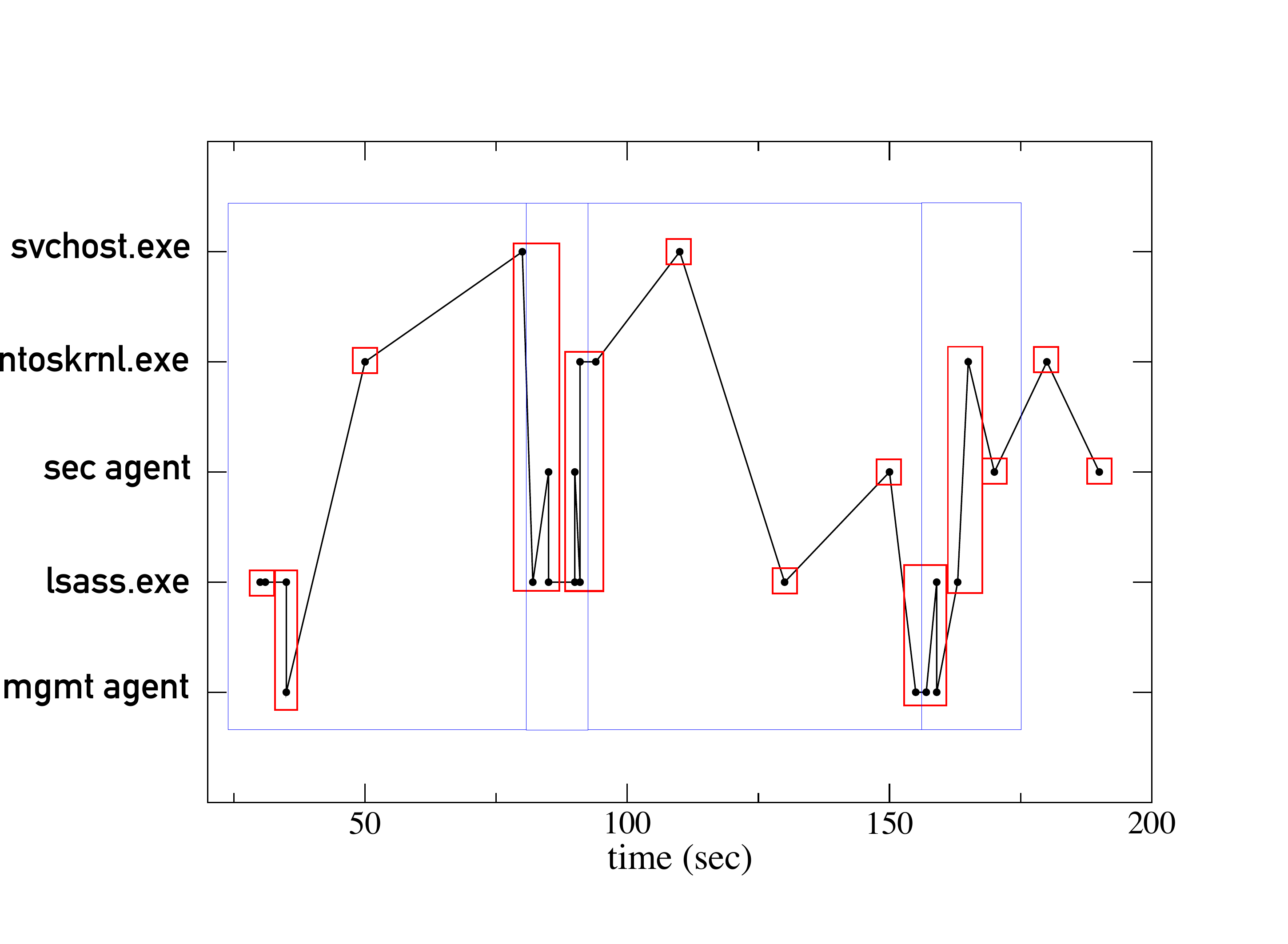}
\caption{\footnotesize{Time series of inter-system communications by process type between a single client and a Domain Controller. Process clusters identified via density-based clustering in time are shown in red.  The stide locality frame is shown in blue for comparison.}}
\label{DCszoom}
\end{figure}
Figure \ref{DCszoom} shows what these clusters look like (red boxes) for a sample process time series of connections between a workstation and a Domain Controller.  We include the stide window with $k=6$ for comparison, to show how it rather arbitrarily combines processes that are likely parts of different system functions.  Though not visible in the plot, occasionally two or more processes will execute simultaneously (and so overlap in the time series plot).  For example, the fourth cluster in Figure \ref{DCszoom} consists of the processes
\begin{equation}
\{\texttt{svchost,svchost,lsass,sec,lsass,lsass}\}\nonumber,
\end{equation}
wherein the two \texttt{svchost} processes occur simultaneously but are destined for different ports on the Domain Controller.  

The basis of this methodology is that there is generally something behaviorally relevant about these clusters, both in terms of chronology and timing.  Quantitatively, this suggests we might find the same clusters appearing more than once in the history of a given system, or within the histories of systems with the same role.  Conversely, rare clusters might indicate unusual process behavior related to a possible malicious connection.
\begin{figure}[t]
\centering
\includegraphics[width=3.5in]{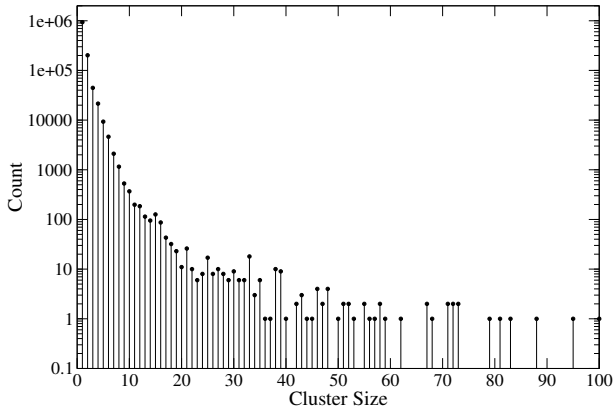}
\caption{\footnotesize{Distribution of process clusters aggregated over the databases of all 125 test systems.}}
\label{clusters_sizes}
\end{figure}

To test this approach, we apply it to the same 125 subjects used previously to test stide.  Each subject's historical database is still organized by peer system, but instead of equal-sized subsequences, the processes are grouped into density-based clusters.  The sizes of these clusters vary: across the full set of 125 test subjects, cluster sizes follow the distribution of Figure \ref{clusters_sizes}.  Most processes are singletons, separated in time from neighboring processes by more than $\epsilon$ seconds.  Interestingly, clusters can become large; for example, several hundred clusters have 10 or more process.   To test a connection between the subject and one of its peers, the new connection data is similarly clustered and compared against the historical process clusters of the other systems with the peer's role.  Like stide, we count the number of matches and compute the $z$-score of each test sample, where here samples are process clusters instead of fixed-length subsequences.
\begin{figure}[t]
\centering
\includegraphics[width=3.5in]{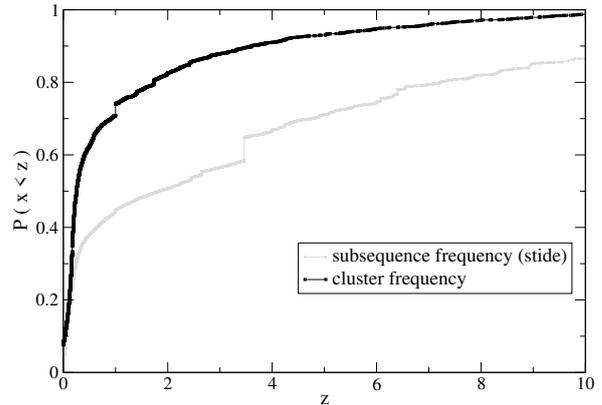}
\caption{\footnotesize{Cumulative probability distribution of test sample $z$-scores resulting from comparison of density-based process clusters, aggregated over all 125 test subjects for one 24-hour test period (black).  The results of stide, discussed previously, are provided for comparison (gray).}}
\label{stide_clusters}
\end{figure}

These results are shown in Figure \ref{stide_clusters}.  There is dramatic improvement over stide; for example, only 10\% of clusters have $z$-scores exceeding 4.  By clustering processes in time, we are doing a better job of capturing the regularities of the process dynamics for each system role.  In essence, density-based process clusters are more {\it meaningful} than  subsequences with fixed locality frame sizes.  Another advantage of using clusters over subsequences is that there are a factor of 10 fewer of them, and so not only is there a smaller percentage of samples lying beyond a given value of $z$, but also fewer in absolute number. 

\begin{figure*}[h]
\centering
\includegraphics[width=5.0in]{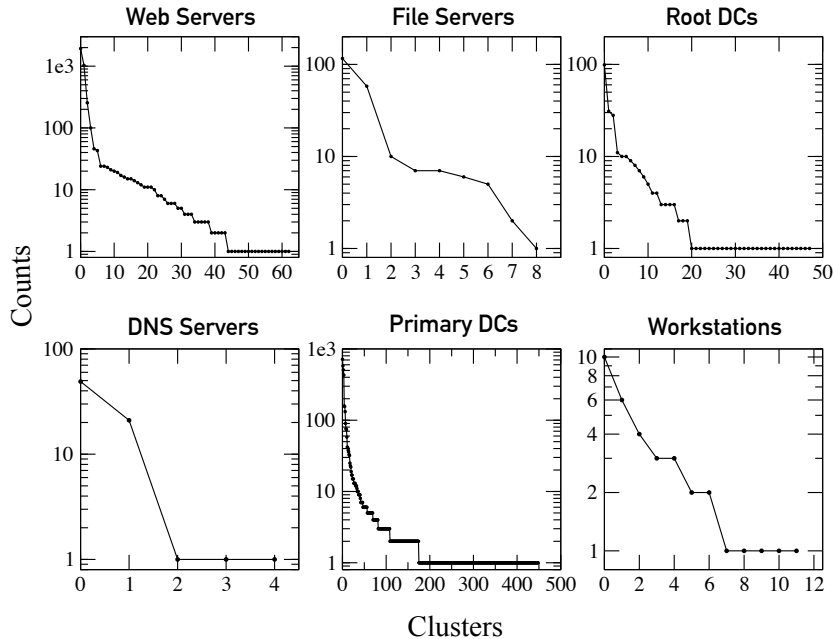}
\caption{\footnotesize{Counts of distinct process clusters for different roles among a single test subject's peers, ranked from high to low.}}
\label{procs_dists}
\end{figure*}

We now look at how these process clusters are distributed for different system roles.  In Figure \ref{procs_dists} we show the process cluster distributions of six roles: internal web services, file servers, root Domain Controllers, DNS servers, primary Domain Controllers, and client workstations.    
Along the $x$-axis of each subplot are the distinct process clusters, like \{\texttt{lsass}, \texttt{svchost}\} and \{\texttt{ntoskrnl}, \texttt{svchost}, \texttt{svchost}\}, though here given number labels to conserve space.  The $y$-axis gives their frequency of occurrence in the historical database of the role. Most roles have many rare clusters; for example, 60\% (amounting to several hundred) of Primary DC clusters are unique (have a single occurrence in the database).  Conversely, most roles exhibit their process dynamics in terms of only a few types of clusters; for example, for web servers, a single cluster accounts for half of all occurrences.  These examples suggest an interesting claim, that relatively few clusters account for a good majority of all occurrences within a role, and that there are moderately long tales of rare clusters.  The first part of this claim supports the idea that systems with the same role ``act the same'' with respect to processes dynamics, and furthermore that density-based clustering is able to isolate these dynamics in terms of frequently-occurring clusters. But, the second part of the claim contends that there is still considerable variability in the process dynamics of system roles, manifested in a preponderance of rare clusters.  To understand this tension, we now take a closer look at these rare clusters. 

If we look at the file server role, there is a single cluster that occurs only once: it is labeled `8' in the plot, but its true identity is the doublet \{\texttt{ntoskrnl}, \texttt{ntoskrnl}\}.  The process \texttt{ntoskrnl} that comprises it happens to be the most common cluster when it occurs alone (labeled `0' in the plot, it is a singleton cluster accounting for around half of all occurrences).  Now, if the DBSCAN $\epsilon$ parameter, which roughly defines the maximum time separation between processes to be considered part of the same cluster, were increased a small amount this rare doublet would separate into these more common singlets, and there would be no novel process clusters in the file server role.  This observation suggests that, though we have attempted to set $\epsilon$ to an appropriate time separation for data with {\it MinPts} = 2,  process clusters might not map so cleanly onto higher-level system functions; some clusters might simply be associations of unrelated processes executing close together in time.  The correct way to view \{\texttt{ntoskrnl}, \texttt{ntoskrnl}\} is then perhaps not as some rare and significant combination of two sequential processes, but instead as two separate \{\texttt{ntoskrnl}\} processes that just happened to occur in close succession.  We should then consider the doublet \{\texttt{ntoskrnl}, \texttt{ntoskrnl}\} to be essentially {\it as common} as the singleton \{\texttt{ntoskrnl}\} out of which it is built.  We are therefore interested in relating the frequency of a process cluster to the frequency of its sub-clusters.  This is essentially the problem of {\it frequent item-set mining} in transactional databases, which we now describe.  
\subsection{Finding frequent sub-clusters}
The particular method we adopt here is Krimp \cite{krimp1,krimp2,krimp3}, which identifies frequent item-sets in a database as those which maximally compress it.  There are other compression-based methods (see, for example, \cite{gokrimp1,gokrimp2,slim,sqs,ism,squish} and the comprehensive survey \cite{MDLsurvey}) but Krimp is technically and conceptually simple and it applies cleanly to our use-case. Krimp identifies frequent subsets in a database of transactions, and then uses these subsets to build compressed instances of each transaction.  The length of the compressed transaction can then be used to infer its typicality: those transactions with comparatively long lengths are candidate outliers \cite{oddoneout,akoglu}.  
In our application, ``transactions'' are process clusters and we apply Krimp ultimately in order to identify relatively incompressible clusters.  Such incompressible clusters are anomalous, signalling that there is something unusual about the timing, ordering, prevalence, or type of processes that comprise it.  The concept of identifying anomalies by process sequences is by no means married to this particular technique, and we consider several other common approaches from the literature later in this paper.   

Krimp works as follows. Given a database, $D$, and set of models, $\mathcal{M}$, the model $M \in \mathcal{M}$ that minimizes the {\it description length},
\begin{equation}
L(M) + L(D|M),
\end{equation}
is the {\it optimal compressor} of the data.  Krimp is a method for finding an approximation of this optimal compressor.  In the following, we briefly review Krimp using the notation and terminology as presented in \cite{krimp1,krimp2}, and point out a few key modifications to the original implementation needed for our application.  Krimp concerns databases built out of discrete items from a set, $\mathcal{I}$.  A transaction, $t$, is a {\it sequence}\footnote{In \cite{krimp1,krimp2}, $t$ is a {\it set} and so items cannot be repeated and order is irrelevant. These constraints are inappropriate for our application, since the same process can meaningfully occur multiple times in a cluster, and chronology is important. For these reasons we define $t$ to be a sequence.}  drawn from $\mathcal{I}$.  A transaction of length $n$ therefore belongs to the set $\mathcal{I}^n$.  As a sequence, the order of $t$ matters and items within $t$ can be repeated.  In our application items are individual processes, like \texttt{lsass.exe} or \texttt{svchost.exe}, and the set $\mathcal{I}$ is the collection of all such processes.  A transaction, $t$, is then a process cluster. The database, $D$, is the collection of all process clusters over a certain time period between a given system and its peer systems within a given role.  

Krimp seeks to compress the database, $D$, by identifying frequent item-sequences, $X \in \mathcal{I}^n$, appearing in the set of transactions\footnote{Contrast with the {\it itemsets}, $X \subseteq \mathcal{I}$ of \cite{krimp1,krimp2}, which, again, do not consider order or allow for repeated items.}.  The models considered by Krimp are {\it code tables}, $CT$, which are simply lists of these item-sequences along with their encodings.  The optimal compressor is then the code table which leads to the shortest encoding of the database, where the length is computed as 
\begin{equation}
L (D |CT) = \sum_{t \in D} L( t | CT).
\end{equation}
The length of a transaction, $t$, is given by the lengths of the encoded item-sequences that appear in it,
\begin{equation}
L(t|CT) = \sum_{X\in cov(t)} L(X|CT),
\label{Lt}
\end{equation}
where $cov(t)$ is the set of item-sequences appearing in, or {\it covering}, the transaction, $t$.  The item-sequences that cover a given transaction must be disjoint (that is, each item in $t$ must belong to only one item-sequence).  Finally, the length of the item-sequence is where the compression comes in: the encoding is based on the frequency of the item-sequence in the database,
\begin{equation}
L(X|CT) = -\log(P(X|D)) = \frac{usage(X)}{\sum_{X'\in CT}usage(X')}.
\end{equation}
The function $usage(X)$ counts the number of transactions with $X$ in their covers.  Krimp begins with the standard code table, which includes only item-sequences corresponding to individual items, and successively adds composite item-sequences one at a time: if the database encoding length is reduced, the item-sequence is added permanently to the code table; otherwise, it is discarded permanently.   In this way, Krimp is a greedy algorithm that works to identify the collection of item-sequences that best cover the transactions in the database, that is, that lead to a shortest encoding of the entire database.  

In our application, item-sequences are sub-clusters: sub-sequences of consecutive processes that make up larger clusters.  We wish to apply Krimp to identify those common sub-clusters that entail an optimal compression of the collection of process clusters.  Clusters that fail to compress well in comparison with the bulk of the collection are considered via this method to be anomalous.  

\begin{figure}[h]
\centering
\includegraphics[width=3.5in]{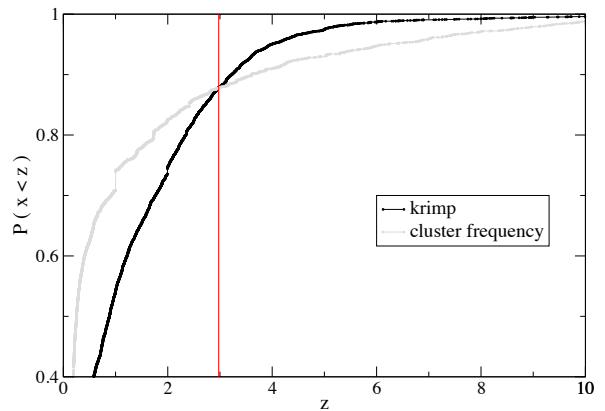}
\caption{\footnotesize{Cumulative probability distribution of test sample $z$-scores resulting from Krimp (black) and a clusters frequencies (gray), aggregated over all 125 test systems for one 24-hour test period.  The red vertical line marks where the Krimp model becomes preferable (smaller probability mass beyond the given $z$-score).}}
\label{krimp_freqs}
\end{figure}

We now apply Krimp to our 125 test subjects, with results presented in Figure \ref{krimp_freqs}.   
The larger cumulative probability at small $z$ for the method based on cluster frequencies reveals that the distributions of $z$-scores under this method are more centralized than under Krimp; however, Krimp has skinnier tales.  This is evident in the crossing of the distributions at around $z = 3$: Krimp has less probability mass below a given $z$ for $z \gtrsim 3$.  Since anomaly detection pertains to the tails of a distribution (large $z$), this is an important finding.  For example, if we set an anomaly threshold at $z=4$, Krimp labels 5\% of the test samples anomalous versus 10\% using cluster frequency analysis.  With generally a factor of two improvement, compression based on frequent item-set mining, as demonstrated by Krimp, appears to resolve additional structure within density-based process clusters useful for understanding process dynamics.  

\begin{figure*}[ht]
\centering
\includegraphics[width=5.0in]{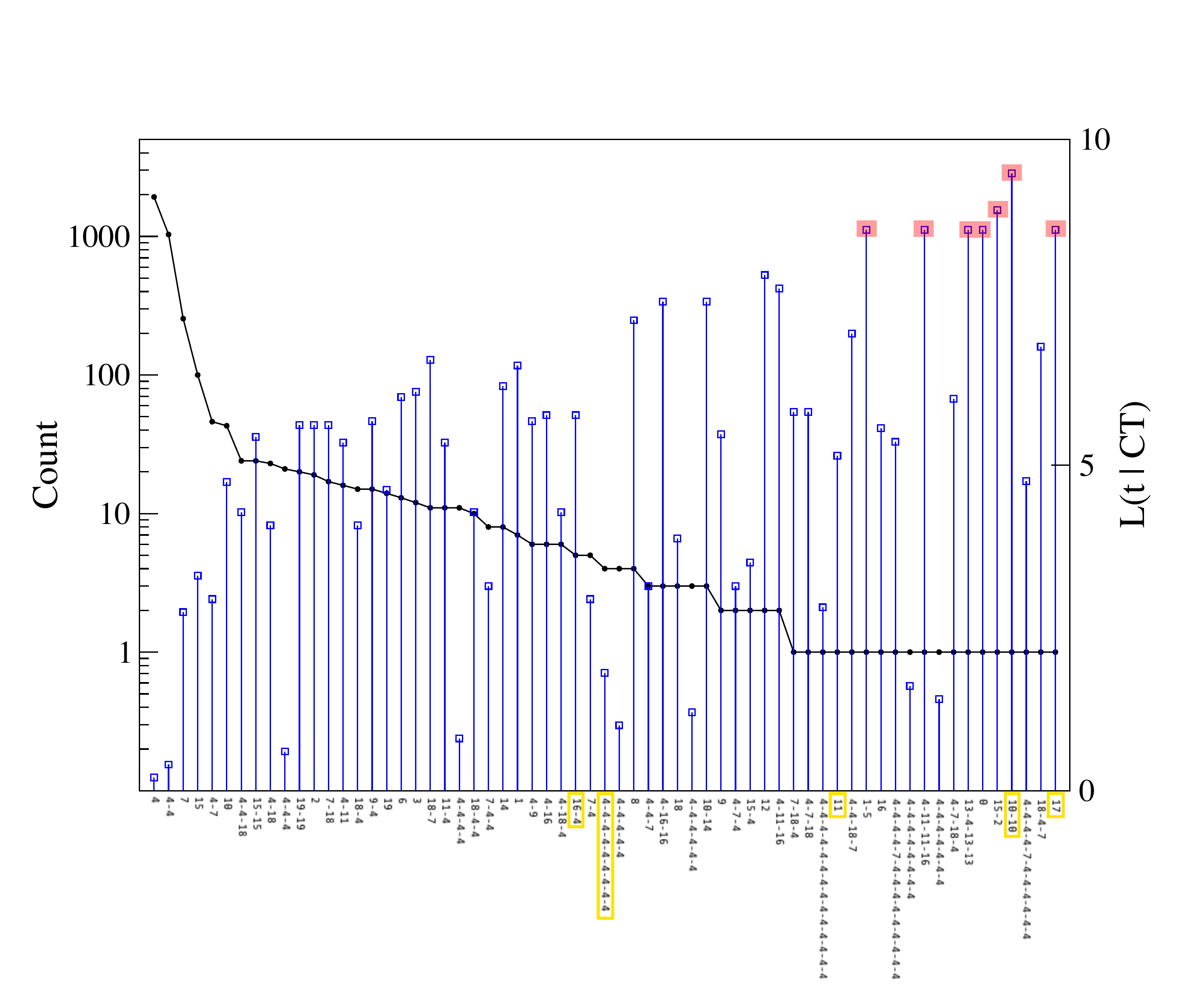}
\caption{\footnotesize{Count (black dots) and encoded length (blue boxes), $L(t|CT)$, of each process cluster found for the web server role over the historical period considered in this analysis. Individual processes are numbered 0 through 19, and the process clusters are labeled along the $x$-axis. Red-highlighted boxes are process clusters deemed anomalous using the threshold discussed in the text.  Process clusters highlighted in yellow are referenced in the text.}}
\label{web_comp}
\end{figure*}

To gain some intuition for how things improve with Krimp, we revisit the results of Figure \ref{procs_dists} showing the frequencies of process clusters for different system roles.   In Figure \ref{web_comp}, we plot the length of the encoded process cluster, $L(t|CT)$, versus its frequency for the web server role as an example.

Suppose that we wish to identify only novel clusters as anomalies. In this case, there are 19 clusters that appear only once in the web server role and are anomalous according to cluster frequency.  In looking at $L(t|CT)$, we see that the encoding length does not have a simple dependence on the cluster's frequency.  This has the welcome effect that most of these 19 clusters have relatively small encodings and therefore look normal.  We set the detection threshold of Krimp using the $z$-score of the novel clusters from the cluster frequency distribution (corresponding to an encoding length $L(t|CT) = 8.61$), and find that only 7 of the 19 singleton clusters are deemed anomalous, a reduction in false alarm rate of more than a factor of two.  This is consistent with the more general performance comparison undertaken above (cf. Figure \ref{krimp_freqs}.) 

To understand this, notice that process \texttt{4} in Figure \ref{web_comp} is the most prevalent process cluster and hence will have the shortest encoding as per Eq. (\ref{Lt}). As a result, clusters built from process \texttt{4} will likewise have short encodings: for example, the large cluster of nine process \texttt{4}'s (highlighted in the figure) has a shorter encoding than {\it smaller} clusters that happen to appear even {\it more} often, like the cluster $\{\texttt{16},\texttt{4}\}$ (also highlighted).  Another property of Krimp is that singleton clusters that appear infrequently can still have short encodings if the process finds itself in other, larger clusters (the converse of the effect seen with process \texttt{4}).  For example, process \texttt{11} occurs only once as a singleton cluster, but has a shorter encoding than other once-appearing singletons, like process \texttt{17} (both highlighted).  As one looks through the processes listed along the $x$-axis, process \texttt{11} appears repeatedly in other clusters, with the result that its encoding is shortened relative to processes like \texttt{17} which only appears as a singleton cluster.

Finally, we note a somewhat counter-intuitive result of Krimp: how is it that certain infrequent process clusters, like cluster \texttt{10-10} (highlighted in Figure \ref{web_comp}), have longer encodings than they would if treated as a single, novel item-sequence?  Since \texttt{10-10} appears only once, we might expect that it should have an encoding length no longer than other once-appearing clusters, like process \texttt{0}, but in fact we find that it is a little longer.  This is because the item-sequence $\{\texttt{10}\}$ has already been included as part of the standard code table, and it is more economical overall to simply build \texttt{10-10} out of this item-sequence then to introduce the item-sequence $\{\texttt{10-10}\}$. Krimp is the consummate utilitarian, doing what is best for the entire database, sometimes at the cost of longer encodings for a few transactions. 

As we close this section, we summarize our journey: we started with stide, which identified anomalies in $k$-length subsequences according to their prevalence in the database.  It performed poorly because temporal information was ignored, and consecutive processes were generally unrelated at the system function level.  To recover this temporal structure, with the hopes of grouping together processes participating in the same higher-level functions, we applied density-based clustering in time.  The frequency distribution of these clusters was considerably more centralized than that of stide, indicating that the clustering allows us to better model the regularities of inter-system processes within a given system role.  This method was also imperfect, as happenstance occasionally prevailed to cluster together unrelated processes, contributing to the preponderance of rare clusters.  To address this problem, we performed frequent item-set mining to the database of clusters to identify common substructures within these clusters, and found that the encoding length of each cluster was revealed to be a better indicator of novelty than its frequency.  
\section{Results and comparison with other methods}
The analysis up to now has focused on how well Krimp is able to model the normal process activity within a given system role.  In particular, we studied Krimp's sensitivity by analyzing the distribution of $z$-scores of encoding lengths, $L(t|CT)$, for each class.  

We now perform an experiment to study how well Krimp identifies normal process clusters as normal and anomalous process clusters as anomalies in practice when deployed on an operational network.  We also compare Krimp against several other methods from the recent literature.

To perform this test, we use the same 125 subjects as before with the same histories and 24-hour test data.  After resolving each subject's peers into roles, to each role we add a single process cluster chosen at random from a different, randomly selected role.  Since the added cluster is not from the role, it is generally anomalous (though, in practice, different roles can have the same process clusters, particularly those involving very common processes).  We test Krimp's ability to spot these anomalous processes while recognizing those rightly belonging to each role as normal.

We compare Krimp's performance with a variety of other approaches from the recent literature. We select Interesting Sequence Miner (ISM) \cite{ism}, which is a compression-based frequent item-set mining algorithm that shares a genealogy with Krimp, in order to test Krimp against a similar, slightly more sophisticated approach.  We also compare against three unrelated methods: frequent pattern outlier factor (FPOF) \cite{fpof}; a categorical data version of the local outlier factor, called $\kappa$-LOF \cite{kappalof}; and common-neighbor-based outlier factor (CNB) \cite{cnb}. Though based on frequent patterns, FPOF is non-compressive and its outlier factor is determined by a support threshold; the $\kappa$-LOF and CNB methods are density- and distance-based, respectively.  Together, the four comparison models are of fundamentally different types and so offer a glimpse of how conceptually distinct outlier factors tackle our problem.   

There are many dozens more methods in the relevant areas of anomaly detection in categorical data \cite{Taha}, frequent pattern mining \cite{Aggarwal2014}, and discrete sequences \cite{Chandola,Domingues}, and it would be prohibitive to consider them all.  We have selected models for comparison that are directly applicable to our problem (that can accommodate sets or sequences of varying sizes, even those containing a single element) and those with publicly available software or algorithms reasonably easy to code from scratch.   For example, while the field of association rule mining offers several approaches useful for anomaly detection in discrete categorical sequences, it is not clear how best to adapt these methods to address test sequences with only a single element, a very common occurrence in our use-case.  We now briefly describe each method.

The ISM is a generative model that builds a database of sequences out of a small set of interesting subsequences, $\mathcal{I}$.  The database is generated by randomly interleaving these subsequences with different multiplicities.  ISM learns the probability distribution, $\Pi$, of interesting subsequences as follows: like Krimp, it begins with only singleton subsequences (the {\it standard code table} in Krimp parlance), and then it iteratively adds candidate subsequences to $\mathcal{I}$ and performs expectation-maximization to optimize the parameters of the distribution.  When this process completes, a sequence, $X$, from the database can be encoded via Shannon's theorem with approximately $-\log_2 p(X | \Pi,\mathcal{I})$ bits.  As a compression-based frequent item-sequence miner, ISM is similar to Krimp; however, Krimp uses a greedy heuristic to generate its code table, and its construction rules are simpler (multiplicity and concatenation, with no interleaving). 

The frequent pattern based outlier detection of \cite{fpof}  mines frequent itemsets in a database by directly appealing to their {\it support}: the support of an itemset, $X$, is the percentage of transactions, $t$, in the database for which $X \subseteq t$.  Given a set of items, $\mathcal{I}$, the set of {\it frequent patterns} (FPS) are those sets $X \subseteq \mathcal{I}$ with at least $s_0$ support.  The frequent pattern outlier factor (FPOF) of $t$ is then defined,
\begin{equation}
{\rm FPOF}(t) = \frac{\sum_{X \subseteq t, X \in {\rm FPS}} {\rm supp}(X)}{|{\rm FPS}|},
\end{equation}  
where ${\rm supp}(X)$ is the support of itemset $X$.  The FPOF of a transaction is simply the percentage of frequent patterns appearing in it, and so small scores suggest anomalies.  As done with Krimp, we make obvious adjustments to adapt this method to apply to sequences instead of sets. The minimal support, $s_0$, is a free parameter that can be tuned to performance requirements.

The $\kappa$-LOF was conceived as a categorical version of the well-known local outlier factor \cite{kappalof} originally devised for numerical data.  The method is given an undirected graph representation, with each transaction in the database a vertex, and edges connecting vertices with a weight proportional to their similarity.  The notion of similarity employed here is based on graph walks, where a $\kappa$-{\it walk} between two vertices $t$ and $t'$ is any sequence of $\kappa$ edges starting a $t$ and ending at $t'$.  The {\it similarity of $\kappa$-walks} between two transactions $t$ and $t'$ is $s^{\kappa}(t,t') = \sum_{t'',t'} w(t'',t') s^{\kappa-1}(t,t'')$, where $w(t'',t')$ is the {\it weight} of the edge between vertices $t''$ and $t'$, and $s^0 = 1$.  In \cite{kappalof}, the weight is defined as the number of common categorical elements between transactions $t''$ and $t'$; since our transactions are in general different lengths, we define the weight as the number of matches relative to the length of the longer transaction.  The {\it accumulated similarity} between $t$ and $t'$ is $S^\kappa(t,t') = \sum_{i=1}^\kappa s^i(t,t')$, and this quantity is used to define the outlier factor,
\begin{equation}
\kappa{\rm -LOF}(t) = \frac{{\rm avg}\{S^\kappa(t,t') | S^\kappa (t,t') > 0\}}{S^\kappa (t,t)}
\end{equation}
for vertices $t'$ reachable from $t$ within $\kappa$-walks.  The denominator is the accumulated similarity of closed walks (those that start and end on $t$) which acts to measure the similarity of $t$ with its local neighborhood; meanwhile, the numerator measures the average similarity of $t$ with vertices further away, up to a distance $\kappa$.  If $\kappa$-LOF$(t)$ is small, then $t$ is more similar to its immediate neighbors than these others, and is not considered an outlier.  Conversely, large $\kappa$-LOF$(t)$ indicates that $t$ is in a neighborhood with vertices more similar to each other than they are to $t$.  This makes $t$ anomalous according to the paradigm of the density-based LOF. One important aspect of this method is that no account is taken of how frequently a transaction appears in the database.  This would need to be incorporated as a vertex attribute of some sort, but does not appear to be considered in \cite{kappalof}.  As we will see, this degrades performance on our database of clusters, for which frequency of a cluster (or its sub-clusters) is an essential aspect of its novelty.

Finally, the CNB method computes the ``distance'' between a transaction $t$ and all others, and defines an outlier factor based on the distance to $t$'s $k^{th}$-nearest neighbor.  The method begins with a notion of similarity, defined in \cite{cnb} as the number matches between equal-length transactions.  Because our transactions are of variable length, we base similarity on common subsequences: the similarity between transactions $t$ and $t'$ is defined as the average length of the largest closed common subsequences relative to the length of the longer transaction.  For example, for $t = (3,4,1,2)$ and $t' = (1,2,3,4)$, the two largest closed common subsequences are $(1,2)$ and $(3,4)$, with an average length of 2.  Following \cite{cnb}, the similarity measure is used to construct the neighbor set of $t$, $NS(t)$, including all transactions $t'$ with a similarity to $t$ greater than some threshold, $\theta$.  The common neighbor set, CNS, between two transactions $t$ and $t'$ is then defined
\begin{equation}
{\rm CNS}(t,t',\theta) = NS(t,\theta) \cap NS(t',\theta).
\end{equation}
The distance between $t$ and $t'$ is 
\begin{equation}
d(t,t') = 1-\frac{\log_2|{\rm CNS}(t,t',\theta)|}{\log_2|\mathcal{D}|},
\end{equation}
where $\mathcal{D}$ is the database and vertical bars denote cardinality.  This distance has a simple interpretation: two transactions are close-together if they have many neighbors in common; in our problem, two process clusters are closer together the more sub-clusters they have in common.  The outlier factor is then sum of the distances between $t$ and its $k$-nearest neighbors. The CNB method includes two free parameters: the similarity threshold, $\theta$, and $k$.  Like $\kappa$-LOF, this method also does not take into account transaction frequency when assessing novelty. 

\begin{figure}[h]
\centering
\includegraphics[width=3.5in]{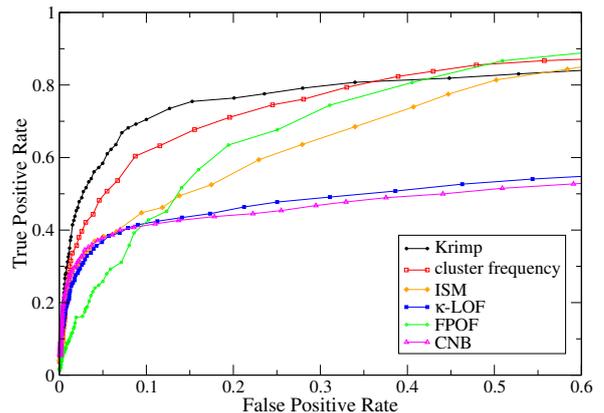}
\caption{\footnotesize{ROC curves for Krimp, cluster frequency, interesting sequence miner (ISM), $\kappa$-local outlier factor, frequent pattern outlier factor (FPOF), and common neighbor-based outlier factor (CND).}}
\label{rocs}
\end{figure}

We plot the receiver operating characteristic (ROC) curves of each model in Figure \ref{rocs}.  For methods with free parameters (FPOF, $\kappa$-LOF, and CNB), we performed a grid search and report results for the model with the largest area under curve (AUC).   
FPOF is sensitive to the support threshold, $s_0$: as $s_0$ is increased, novelty becomes more commonplace as only the most frequent itemsets are included in FPS.  We find that AUC is greatest with $s_0 = 3$.  The $\kappa$-LOF has the greatest time complexity, $\mathcal{O}(n^2(q+\kappa))$ for $n$ transactions and $q$ within $\kappa$-walks of transaction $t$: with $\kappa > 2$, the time-performance trade-off tips heavily against this method.  For CND, AUC is best for $\theta = 0.25$ and $k = 2$.

In all, Krimp performs best with AUC $=0.8$, followed by the cluster frequency method discussed in the last section.  Krimp's strengths are its emphasis on frequent item-sequences (and its use of compression rather than a threshold criterion as in FPOF), and its relatively simple mining heuristics; the interleaving used by ISM to build sequences is evidently not a useful symmetry of the higher-level system functions to which process clusters correspond.      

\section{Conclusions}
In this paper, we present an unsupervised framework for lateral movement detection on enterprise networks.  The framework comprises two detection techniques aimed at different aspects of the lateral movement discovery problem.  These techniques make essential use of the concept of system {\it role}.  Over the course of normal operations, systems tend to make connections to remote systems of a small and stable set of roles, such that connections to systems of novel roles can be identified and investigated as potential lateral movement.  Furthermore, the processes that underlie these connections follow temporal patterns based on the roles of the systems involved in the connection, such that deviations from these expected patterns signal anomalous activities.   

These methods show promise on large, operational enterprise networks. Role-based anomaly detection plateaus to a stable number of alerts collected over a set of watched systems: in our test sample of 125 subjects, the method averages 25 alerts daily.  Role-process-based anomaly detection also performs well, with an AUC of around 0.8 that outperforms alternative methods of anomaly detection.  Big data platforms can be leveraged to perform this analysis across a large portion of the network: if the number of false alarms gets large, alerts can be incorporated into a correlation process where they are considered along with other alerts and indicators.   

This work might be extended in a number of ways.  As presented here, it is entirely unsupervised; however, it might be possible to incorporate one-class learning into role-based anomaly detection by training a model on each role just once.  There would be a single model for each role across the entire watch list (rather than roles being found anew for each subject during each test period).  By foregoing the unsupervised clustering step, this method could be considerably sped up.  The model could also be matured and updated over time, potentially improving accuracy over unsupervised methods.  

As presented here, role-process-based anomaly detection requires that Krimp be run anew each test period; this is time consuming for connection-heavy systems like Domain Controllers and file servers.  An iterative process that more efficiently updates established code tables with new transactions would be a valuable speed up.   

Role-process-based anomaly detection could be extended by incorporating additional traffic characteristics, like bytes transferred per connection.  Thus, in addition to process patterns, normal system functions could be characterized by typical data transfer rates.  This would require correlating a data source with this traffic information, like Netflow, with process data.   

Lateral movement via authorized means is a common tactic of advanced threats, and its detection remains a great challenge to the organizations they target.  Cyber defense must move beyond rule sets and signature-based detection in order to resist these threats, and we hope this framework, which leverages common data and uses standard algorithms, can be incorporated into the defense-in-depth of the vulnerable networks in the cross-hairs of the relentless and worthy adversary. 
\bibliography{powell}

\end{document}